\documentstyle[preprint,aps]{revtex}
\tighten
\draft
\widetext
\input epsf
\preprint{YITP-SB-02-25}
\bigskip
\bigskip

\def\b{\overline}
\def\t{\widetilde}

\begin{document}
\title{Toward Solving the Cosmological Constant Problem?}
\medskip
\author{Olindo Corradini$^1$\footnote{E-mail: corradini@bo.infn.it},
Alberto Iglesias$^2$\footnote{E-mail: iglesias@insti.physics.sunysb.edu} and
Zurab Kakushadze$^3$\footnote{E-mail: zurab.kakushadze@rbccm.com}}
\bigskip
\address{$^1$Dipartimento di Fisica, Universit\`a di Bologna and INFN, 
Sezione di Bologna\\
via Irnerio 46, I-40126 Bologna, Italy\\
$^2$C.N. Yang Institute for Theoretical Physics\\ 
State University of New York, Stony Brook, NY 11794\\
$^3$Royal Bank of Canada Dominion Securities Corporation\footnote{This address
is used by the corresponding author for no purpose other than to indicate his
professional affiliation as is customary in scientific publications. In 
particular, the contents of this paper are limited to Theoretical Physics, have
no relation whatsoever to the Securities Industry, do not have any value for
making investment or any other financial decisions, and in no way represent
views of the Royal Bank of Canada or any of its other affiliates.}\\
1 Liberty Plaza, 165 Broadway, New York, NY 10006}
\date{December 10, 2002}
\bigskip
\medskip
\maketitle

\begin{abstract} 
{}We discuss the cosmological constant problem in the context of 
higher codimension brane
world scenarios with infinite-volume extra dimensions. In particular,
by adding higher curvature terms in the bulk action we are able to find
smooth solutions with the property that the 4-dimensional part of the
brane world-volume is flat for a range of positive values of the brane
tension.
\end{abstract}

\section{Introduction and Summary}

{}As was proposed in \cite{DG}, 
one can reproduce four-dimensional gravity
on a 3-brane in 6 or higher dimensional infinite-volume bulk if one includes
the Einstein-Hilbert term on the brane (the latter is induced at the quantum 
level if the brane matter is non-conformal). Gravity then is almost
completely localized on the brane with only ultra-light modes penetrating into
the bulk, so that gravity is four-dimensional at distance scales up to
an ultra-large cross-over scale $r_c$ (beyond which gravity becomes higher 
dimensional), which can be larger than the present Hubble size.
In particular, this is the case for codimension-2 and higher 
tensionless branes \cite{DG} as well as for codimension-2 non-zero tension
branes \cite{CIKL}. This is also expected to be the case for codimension-3
and higher non-zero tension branes \cite{orient,DGHS}.

{}A careful analysis of linearized gravity in such backgrounds requires
smoothing out higher codimension singularities \cite{DG,orient}. This is
already the case for tensionless branes, where the background is non-singular 
(in fact, it is flat\footnote{More precisely, in the absence of higher 
curvature terms in the bulk the latter can be Ricci-flat (here and in the
following we focus on the solutions with vanishing brane cosmological 
constant).}), 
but the graviton propagator is singular. However, as 
was pointed out in \cite{DGHS}, these singularities in the graviton 
propagator are cured once we include higher curvature terms on the {\em brane}.
These terms are induced at the quantum level along with the Einstein-Hilbert 
term, and provide an ultra-violet cut-off for the graviton propagator.
In the presence of these terms, to reproduce four-dimensional gravity up
to the present Hubble size, we need to assume that the bulk Planck scale 
$M_P{\ \lower-1.2pt\vbox{\hbox{\rlap{$<$}\lower5pt\vbox{\hbox{$\sim$}}}}\ }
(1~{\rm mm})^{-1}$ regardless of the codimension of the brane
(assuming that it is 2 or higher) \cite{DGHS}.

{}In the case of non-zero tension branes, however, the situation is more 
complicated as in the case of $\delta$-function-like branes
the background itself becomes singular (for phenomenologically interesting
non-BPS branes on which we focus in this paper). More precisely, here we focus
on solutions with vanishing cosmological constant on the brane. Then in the
codimension-2 case the singularity is very mild as in the extra 2 dimensions
the background has the form of a wedge with a deficit angle, so the 
singularity is a simple conical one \cite{CIKL}. As was discussed in 
\cite{orient}, this singularity can be consistently smoothed out. In this 
case gravity on the brane was analyzed in \cite{CIKL,orient}, 
where it was found that the behavior of
gravity is essentially unmodified compared with the tensionless brane cases.
In particular, gravity is almost completely localized with only ultra-light
modes penetrating into the bulk, and four-dimensional gravity on the brane
can be reproduced up to distance scales of order of the current Hubble
size. As to the brane tension, in codimension-2 solutions with vanishing
brane cosmological constant it can take continuous values in the range
bounded below by zero and bounded above by its critical value \cite{CIKL}, 
which is of order $M_P^4
{\ \lower-1.2pt\vbox{\hbox{\rlap{$<$}\lower5pt\vbox{\hbox{$\sim$}}}}\ } 
(1~{\rm mm})^{-4}$ (for a 3-brane). Since this does not improve the current
experimental bound on the 4-dimensional cosmological constant, we are
prompted to consider codimension-3 and higher solutions.  

{}In codimension-3 and higher cases with non-zero tension branes
the singularities in the 
background itself are more severe. Actually, there are two types of
singularities we must distinguish here. Thus, we can have a singularity
at the origin $r=0$ of the extra space, that is, at the location of the brane.
These singularities can sometimes be simple coordinate singularities, in which
case they are harmless. On the other hand, true $r=0$ singularities can be
smoothed out using a 
procedure discussed in \cite{orient}. This procedure goes as follows. Consider
a codimension-$d$ ($d>2$)
$\delta$-function-like source brane in $D$-dimensional bulk.
Let the bulk action simply be the $D$-dimensional Einstein-Hilbert action,
while on the brane we have the induced $(D-d)$-dimensional Einstein-Hilbert 
term as well as the cosmological term corresponding to the brane tension.
As we have already mentioned, the background in this case is singular 
\cite{Gregory}. One way to smooth out such a singularity is to replace the
$(D-d)$-dimensional world-volume of the brane by its product with a 
$d$-dimensional ball ${\bf B}_d$ of some non-zero radius $\epsilon$. 
As was pointed out in \cite{orient}, in this case already for a tensionless 
brane the gravitational modes on the brane contain an infinite tower
of tachyonic modes. This can be circumvented by considering a partial smoothing
out where one replaces the  
$(D-d)$-dimensional world-volume of the brane by its product with a 
$(d-1)$-sphere ${\bf S}^{d-1}$ of radius $\epsilon$ \cite{orient}. 
As was pointed out in \cite{orient}, this suffices for 
smoothing out higher codimension singularities in the graviton propagator
as in the codimension-1 case the propagator is non-singular \cite{DGP}. 
Moreover,
in the case of tensionless branes as well as in the case of a non-zero
tension codimension-2 brane we then have only one tachyonic mode 
(with ultra-low $-m^2$) which is expected to be an
artifact of not including non-local operators on the brane. As to the 
background itself, this procedure does smooth out $r=0$ singularities.
However, not surprisingly, it does not suffice for smoothing out the second
type of singularities \cite{Higher} (such singularities are absent in the
codimension-2 case but are present in higher codimension cases with
non-zero brane tension), which we discuss next.

{}Thus, in codimension-3 and higher cases with vanishing brane cosmological
constant we can have singularities at $r=r_*$ with some non-vanishing $r_*$,
which depends on the brane tension
\cite{Gregory}. The aforementioned smoothing out procedure does not cure
such singularities. In \cite{Higher} it was suggested that higher curvature
terms in the {\em bulk} might be responsible for curing such singularities.
The purpose of this paper is precisely to study effects of higher curvature
terms in the bulk on such singularities. 

{}Not surprisingly, studying such backgrounds in the presence of higher 
curvature terms in the bulk is rather non-trivial. To make the problem more
tractable, we focus on a special kind of higher curvature terms in the bulk.
In particular, we consider adding the second-order (in curvature)
Gauss-Bonnet combination in the bulk. This Gauss-Bonnet combination is an
Euler invariant in four dimensions. In higher dimensions, however, it has a
non-trivial effect on the equations of motion if the background is not flat.
A simplifying feature of the Gauss-Bonnet combination is that it does not
introduce terms with third and fourth derivatives in the background 
equations of motions, but rather makes them even more non-linear. Albeit 
non-trivial, these equations can in certain cases be analyzed analytically,
so we can get some insight into the effect of higher curvature terms on
the aforementioned $r=r_*$ singularities.

{}The results of our analyses, which we present in the remainder of this paper,
suggest that higher curvature terms in the bulk might indeed cure these
singularities. To simplify our discussion, as far as explicit computations are
concerned, we focus on codimension-3 cases, where we have additional 
simplifications in the equations of motion (higher codimension cases are
{\em not} expected to be qualitatively different). In particular, we explicitly
study examples of: a string in 5D bulk, a membrane in 6D bulk as well as
a 3-brane in 7D bulk. In these cases we argue that
smooth solutions with vanishing brane cosmological constant do exist for
a continuous range of positive brane tension. In the remainder of this paper,
following the terminology adopted in \cite{DGS}, we refer to such solutions
as {\em diluting} solutions. 

{}Before we turn to a detailed description of our solutions, let us describe
the geometry of the aforementioned diluting solution corresponding to a
3-brane in 7D bulk.  
In the extra three dimensions we have a radially symmetric
solution where a 2-sphere is fibered over a semi-infinite line $[r_0,\infty)$.
The space is curved for $r_0\leq r<\epsilon$, at $r=\epsilon$ we have a jump
in the radial derivatives of the warp factors (because $r=\epsilon$ is where
the brane tension is localized), and for $r>\epsilon$ the space is actually
{\em flat}. So an outside observer located at $r>\epsilon$ thinks that the
brane is tensionless, while an observer inside of the sphere, that is, at
$r<\epsilon$ sees the space highly curved. It would take this observer
infinite time to reach the point $r=r_0$, where we have a coordinate
singularity (that is, the corresponding geodesics are complete, so this is
{\em not} a naked singularity but a coordinate one). This is an
important point, in particular, we did not find smooth
solutions where the space would be curved outside but flat inside of the 
sphere. And this is
just as well. Indeed, as was shown in \cite{Higher}, if we have only the
Einstein-Hilbert term in the bulk, then we have no smooth solutions whatsoever
(that is, smoothing out the 3-brane by making it into a 5-brane with two
dimensions curled up into a 2-sphere does not suffice). What is different in
our solutions is that we have added higher curvature terms in the bulk, which
we expected to smooth out some singularities. But serving as an ultra-violet
cut-off higher curvature terms could only possibly smooth out a real
singularity at $r<\epsilon$, not at $r>\epsilon$. And this is precisely what
happens in our solution - the presence of higher curvature terms ensures that
we have only a coordinate singularity at $r<\epsilon$ instead of a real naked
singularity as would be the case had we included only the Einstein-Hilbert 
term.

\section{Setup}

{}The brane world model we study in this paper is described 
by the following action:
\begin{eqnarray}
 S=&&{\widetilde M}^{D-3}_P \int_{\Sigma} d^{D-1}x~
 \sqrt{-{\widetilde G}}\left[{\widetilde R}-{\widetilde\Lambda}\right]
 +\nonumber\\
 &&{M}^{D-2}_P \int d^{D}x~
 \sqrt{-{G}}~\left[{R}~+\xi\biggl(R^2-4R_{MN}^2+R_{MNPR}^2\biggl)\right].
 \label{action}
\end{eqnarray}
Here $M_P$ is the (reduced) $D-$dimensional Planck mass, while 
${\widetilde M}_P$
is the (reduced) $(D-1)$-dimensional Planck mass; $\Sigma$ is a 
source brane, whose geometry is given by 
the product ${\bf R}^{D-d-1,1}\times {\bf S}^{d-1}_\epsilon$, where 
${\bf R}^{D-d-1,1}$ is the $(D-d)$-dimensional Minkowski space, and 
${\bf S}^{d-1}_\epsilon$ is a $(d-1)$-sphere of radius $\epsilon$ (in the
following we will assume that $d\geq 3$).
The quantity ${\widetilde M}^{D-1}_P {\widetilde \Lambda}$ plays the role of 
the tension of the brane $\Sigma$. Also, 
\begin{equation}
 {\widetilde G}_{mn}\equiv
 {\delta_m}^M{\delta_n}^N G_{MN}\Big|_\Sigma~,
\end{equation} 
where $x^m$ are the $(D-1)$ coordinates along the brane (the $D$-dimensional
coordinates are given by $x^M=(x^m,r)$, where $r\geq 0$ is a non-compact
radial coordinate transverse to the brane, and the signature of the 
$D$-dimensional metric is $(-,+,\dots,+)$); finally, the $(D-1)$-dimensional
Ricci scalar ${\widetilde R}$ is constructed from the $(D-1)$-dimensional
metric ${\widetilde G}_{\mu\nu}$. In the following we will use the notation
$x^i=(x^\alpha,r)$, where $x^\alpha$ are the $(d-1)$ angular 
coordinates on the sphere ${\bf S}^{d-1}_\epsilon$.
Moreover, the metric for the coordinates $x^i$ will be (conformally) flat:
\begin{equation}
 \delta_{ij}~dx^i dx^j=dr^2+r^2\gamma_{\alpha\beta}~dx^\alpha dx^\beta~,
\end{equation}
where $\gamma_{\alpha\beta}$ is the metric on a unit $(d-1)$-sphere.
Also, we will
denote the $(D-d)$ Minkowski coordinates on ${\bf R}^{D-d-1,1}$ via $x^\mu$
(note that $x^m=(x^\mu,x^\alpha)$).

{}The equations of motion read
\begin{eqnarray}
 &&R_{MN}-\frac{1}{2} G_{MN}R+\xi \left[2\Xi_{MN}-{1\over 2} G_{MN}\, 
 \Xi\right]+
 \nonumber\\
 &&\displaystyle{
 {\sqrt{-{\widetilde G}}\over\sqrt{-G}}}
 {\delta_M}^m {\delta_N}^n \left[{\widetilde R}_{mn}-{1\over 2} 
 {\widetilde G}_{mn}\left({\widetilde R}-
{\widetilde \Lambda}\right)\right]  {\widetilde L}~ \delta(r-\epsilon)=0~,
 \label{EoM4}
\end{eqnarray}
where $\Xi\equiv {\Xi_M}^M$, and
\begin{eqnarray}
 &&\Xi_{MN}\equiv 
 R R_{MN}-2R_{MS}R^S{}_N+R_{MRST}R_N{}^{RST}-2R^{RS}R_{MRNS}~,\\
 &&{\widetilde L}\equiv {\widetilde M}_P^{D-3}/M_P^{D-2}~.
\end{eqnarray}
Here we are interested in solutions with vanishing $(D-d)$-dimensional
cosmological constant, which, at the same time, are radially symmetric in the
extra $d$ dimensions. The corresponding ansatz for the background metric reads:
\begin{equation}
 ds^2=\exp(2A)~\eta_{\mu\nu}~dx^\mu dx^\nu+\exp(2B)~\delta_{ij}~dx^i dx^j~,
\end{equation}
where $A$ and $B$ are
functions of $r$ but are independent of $x^\mu$ and $x^\alpha$. 
We then have (here prime
denotes derivative w.r.t. $r$):
\begin{eqnarray}
 &&{\widetilde R}_{\mu\nu}=0~,\\
 &&{\widetilde R}_{\alpha\beta}=\lambda ~{\widetilde G}_{\alpha\beta}~,\\
 &&{\widetilde R}=(d-1)~\lambda~,\\
 &&R_{\mu\nu}=-\eta_{\mu\nu}e^{2(A-B)}
 \left[A^{\prime\prime} +(d-1){1\over r} A^\prime+
 (D-d)(A^\prime)^2 +(d-2)A^\prime B^\prime\right]~,\\
 &&R_{rr}=-(d-1)\left[B^{\prime\prime}+ {1\over r} B^\prime\right]
 +(D-d)\left[A^\prime B^\prime-
 (A^\prime)^2 -A^{\prime\prime}\right]~,\\
 &&R_{\alpha\beta}=-r^2 \gamma_{\alpha\beta}
 \left[B^{\prime\prime}+(2d-3){1\over r}B^\prime +(d-2)(B^\prime)^2+
 (D-d)A^\prime B^\prime + (D-d){1\over r}A^\prime\right]~,\\
 &&R=-e^{-2B}\Big[2(d-1)B^{\prime\prime} + 2(d-1)^2{1\over r} B^\prime +
 2(D-d)A^{\prime\prime} + 2(D-d)(d-1){1\over r} A^\prime 
 +\nonumber\\
 &&(d-1)(d-2)(B^\prime)^2 +(D-d)(D-d+1)(A^\prime)^2
 +2(D-d)(d-2)A^\prime B^\prime\Big]~. 
\end{eqnarray}
where
\begin{equation}
 \lambda\equiv {{d-2}\over \epsilon^2}e^{-2B(\epsilon)}~.
\end{equation}
The equations of motion then read:
\begin{eqnarray}\label{AB1}
 &&{1\over 2}\b d (\b d-1)(A^\prime)^2+\b d(d-1)N+{1\over 2}(d-1)(d-2)S-
 \nonumber\\
 &&\xi\ {\rm e}^{-2B}\biggl\{{1\over 2}\b d(\b d -1)(\b d-2)(\b d-3)
 (A^\prime)^4
 +2\b d(\b d -1)(d-1)(d-2)N^2+\nonumber\\
 &&{1\over 2}(d-1)(d-2)(d-3)(d-4)S^2
 +2\b d(\b d -1)(\b d-2)(d-1)(A^\prime)^2N
 \nonumber\\
 &&+\b d(\b d -1)(d-1)(d-2)(A^\prime)^2S
 +2\b d(d-1)(d-2)(d-3)NS\biggr\}=0~,\\%[3mm]
 &&{1\over 2}(\b d-1)(\b d-2)(A^\prime)^2+(\b d-1)M+(\b d-1)(d-1)N
+(d-1)P+{1\over 2}(d-1)(d-2)S-\nonumber\\
&&\xi\ {\rm e}^{-2B}\biggl\{{1\over 2}(\b d-1)(\b d-2)(\b d-3)(\b d-4)
(A^\prime)^4+2(\b d-1)(\b d -2)(d-1)(d-2)N^2+
\nonumber\\&&{1\over 2}(d-1)(d-2)(d-3)(d-4)S^2
+2(\b d-1)(\b d-2)(\b d-3)(A^\prime)^2M+
\nonumber\\&&2(\b d-1)(\b d -2)(\b d-3)(d-1)(A^\prime)^2N
+2(\b d-1)(\b d-2)(d-1)(A^\prime)^2P+
\nonumber\\&&(\b d-1)(\b d-2)(d-1)(d-2)(A^\prime)^2S
+4(\b d-1)(\b d-2)(d-1)MN+\nonumber\\
&&2(\b d-1)(d-1)(d-2)MS+4(\b d-1)(d-1)(d-2)NP+\nonumber\\
&&2(\b d-1)(d-1)(d-2)(d-3)NS+2(d-1)(d-2)(d-3)PS\biggr\}+
\nonumber\\&&{1\over 2}{\rm e}^{B}
\biggl[\t\Lambda-(d-1)\lambda\biggr]\t L\delta(r-\epsilon)=0~,\label{AB2}\\
%[3mm]
%
%
%
&&{1\over 2}\b d(\b d-1)(A^\prime)^2+\b dM+\b d(d-2)N
+(d-2)P+{1\over 2}(d-2)(d-3)S-\nonumber\\
&&\xi\ {\rm e}^{-2B}\biggl\{{1\over 2}\b d(\b d-1)(\b d-2)(\b d-3)
(A^\prime)^4+2\b d(\b d -1)(d-2)(d-3)N^2+
\nonumber\\&&{1\over 2}(d-2)(d-3)(d-4)(d-5)S^2
+2\b d(\b d-1)(\b d-2)(A^\prime)^2M+
\nonumber\\&&2\b d(\b d-1)(\b d -2)(d-2)(A^\prime)^2N
+2\b d(\b d-1)(d-2)(A^\prime)^2P+
\nonumber\\&&\b d(\b d-1)(d-2)(d-3)(A^\prime)^2S
+4\b d(\b d-1)(d-2)MN+2\b d(d-2)(d-3)MS+
\nonumber\\&&4\b d(d-2)(d-3)NP
+2\b d(d-2)(d-3)(d-4)NS+2(d-2)(d-3)(d-4)PS\biggr\}+
\nonumber\\&&{1\over 2}{\rm e}^{B}
\biggl[\t\Lambda-(d-3)\lambda\biggr]\t L\delta(r-\epsilon)=0~,\label{AB3}
\end{eqnarray}
where we have defined
\begin{eqnarray}
&&M\equiv A^{\prime\prime}+(A^\prime)^2-A^\prime B^\prime~,\hskip1cm
N\equiv A^\prime \left(B^\prime +{1\over r}\right)~,\nonumber\\
&&P\equiv B^{\prime\prime}+{B^\prime\over r}~,\hskip1cm
S\equiv (B^\prime)^2+2{B^\prime\over r}~,\nonumber\\ 
&&\b d \equiv D-d~.\nonumber
\end{eqnarray}
Above the third equation is the $(\alpha\beta)$ equation, 
the second equation is the
$(\mu\nu)$ equation, while the first equation is the $(rr)$ equation.
Note that the latter 
does not contain second derivatives of $A$ and $B$. Also note that the other
two equations do not contain third and fourth derivatives of $A$ and $B$
(this is a special property of the Gauss-Bonnet combination we mentioned in
Introduction).

{}The equations of motion (\ref{AB1})-(\ref{AB3}) are highly non-linear and
difficult to solve in the general case. However, in the $d=4$ and especially
$d=3$ cases various (but not all) 
terms proportional to $\xi$ vanish. This is due
to the fact that the Gauss-Bonnet combination is an Euler invariant in four
dimensions. To make our task more tractable, from now on we will focus on
the codimension-3 case ($d=3$). We do {\em not} 
expect higher codimension cases to
be qualitatively different.

{}Not only is the complexity of the above equations of motion sensitive to
the value of $d$, but also to the value of ${\overline d}$. In particular,
we have substantial simplifications in the cases of ${\overline d}=2$ and 
${\overline d}=3$ corresponding to the non-compact part of the brane $\Sigma$
being a string respectively a membrane. Note that these simplifications are
specific to the Gauss-Bonnet combination. In particular, if we set the 
Gauss-Bonnet coupling $\xi$ to zero (that is, if we keep only the 
Einstein-Hilbert term in the bulk), there is nothing special about the 
${\overline d}=2,3$ cases. This suggest that our conclusions derived from
explicit analytical computations for the 
${\overline d}=2,3$ cases can be expected
to hold in ${\overline d}\geq 4$ cases as well (in particular, in the case
of a 3-brane in 7D and, as we mentioned above, even higher dimensional bulk). 
In fact, our analytical and numerical results in the case of a 
3-brane in 7D bulk confirm this expectation.  

\section{No Einstein-Hilbert Term in the Bulk}

{}To warm up and get a feeling for whether we can expect to find 
smooth solutions in the case where we have the Einstein-Hilbert as well as
the Gauss-Bonnet terms in the bulk, in this section we will discuss a
somewhat simpler (and more of a toy) problem. In particular, we will study the
case where we have no Einstein-Hilbert term in the bulk but only the 
Gauss-Bonnet term. This simplifies the equations of motion enough so that we 
can solve them analytically. In the next section we will come back to our
original problem where we have both the Einstein-Hilbert term as well as the
Gauss-Bonnet term in the bulk.

\subsection{String in 5D Bulk}

{}Let us first consider a string ($\overline 
d=2$) propagating in 5D bulk ($d=3$). 
Our equations of motion then read:
\begin{eqnarray}\label{S-eq1}
&&(A^\prime)^2 \left[3(B^\prime)^2+6{B^\prime\over r}+{2\over r^2}\right]=0~,\\
&& \left[A^{\prime\prime}+(A^\prime)^2-A^\prime B^\prime\right]
\left[(B^\prime)^2+2{B^\prime\over r}\right] +2A^\prime 
\left(B^\prime +{1\over r}\right)\left(B^{\prime\prime}
+{B^\prime\over r}\right)=\nonumber\\
&&{1\over 8\xi}{\rm e}^{3B}(\kappa-2)\lambda\widetilde L
\delta(r-\epsilon)~,\label{S-eq2}\\
&&(A^\prime)^2\left(B^{\prime\prime}+{B^\prime\over r}\right)
+2\left[A^{\prime\prime}+(A^\prime)^2-A^\prime B^\prime\right]
A^\prime \left(B^\prime +{1\over r}\right)=\nonumber\\
&&{\kappa\over 8\xi}{\rm e}^{3B}\lambda\widetilde L
\delta(r-\epsilon)~,\label{S-eq3}
\end{eqnarray}
where we have introduced the notation
\begin{equation}
 \kappa\equiv{\widetilde \Lambda}/\lambda~.
\end{equation}
Let us now solve the above equations of motion.

{}Thus, we have two types of solutions.\\
$\bullet$ Solution A:
\begin{eqnarray}\label{S-sol1A}
&& r<\epsilon:~~~A(r)={\rm constant},~~~B(r)={\rm anything}~,\\
&& r>\epsilon:~~~A(r)={\rm ln}\left({r_A\over r}\right)^{\alpha -1},~~~
B(r)={\rm ln}\left({r_B\over r}\right)^\alpha~.
\label{S-sol2A}
\end{eqnarray}
$\bullet$ Solution B:
\begin{eqnarray}\label{S-sol1B}
&& r<\epsilon:~~~A(r)={\rm ln}\left({r_A\over r}\right)^{\alpha -1},~~~
B(r)={\rm ln}\left({r_B\over r}\right)^\alpha~.
\label{S-sol2B}\\
&& r>\epsilon:~~~A(r)={\rm constant},~~~B(r)={\rm anything}~.
\end{eqnarray}
Here
\begin{equation}
 \alpha=\alpha_\pm = 1\pm {1\over \sqrt 3}~.
\end{equation} 
The jump conditions at $r=\epsilon$ are given by:
\begin{eqnarray}\label{S-jump-a}
&&\left. \left[A^\prime \left((B^\prime )^2+2 {B^\prime\over r}\right)\right] 
\right|_{\epsilon-}^{\epsilon+}
={\kappa -2\over 8\xi}{\tilde L\over \epsilon^2}
{\rm e}^{B(\epsilon)}~,\\
&&\left.\left[(A^\prime)^2\left( B^\prime+ {1\over r}\right)\right]
\right|_{\epsilon-}^{\epsilon+}={\kappa\over 8\xi}{\tilde L\over \epsilon^2}
{\rm e}^{B(\epsilon)}~.
\label{S-jump-b}
\end{eqnarray}
Note that, without losing any solutions parametrically,
we can assume that $B(r)={\rm constant}$
whenever $A(r)={\rm constant}$. The jump conditions then read:
\begin{eqnarray}\label{S-jump-1}
 &&\pm\left.
 \left[A^\prime \left((B^\prime )^2+2 {B^\prime\over r}\right)\right] 
 \right|_{r=\epsilon\pm}
 ={\kappa -2\over 8\xi}{\widetilde L\over \epsilon^2}
 {\rm e}^{B(\epsilon)}~,\\
&&\pm\left.\left[(A^\prime)^2\left( B^\prime+ {1\over r}\right)\right]
\right|_{r=\epsilon\pm}={\kappa\over 8\xi}{\widetilde L\over \epsilon^2}
{\rm e}^{B(\epsilon)}~,
\label{S-jump-2}
\end{eqnarray}
where the upper ($+$) sign corresponds to Solution A, while the lower
($-$) sign corresponds to Solution B. 
We then have:
\begin{eqnarray}
&&
\pm
{\alpha(\alpha-1)(\alpha-2)\over \epsilon}=
 -{\kappa -2\over 8\xi}{\widetilde L}
\left({r_B\over \epsilon}\right)^\alpha~,\\
&&\pm {(\alpha-1)^3\over \epsilon}
=-{\kappa\over 8\xi}{\widetilde L}
\left({r_B\over \epsilon}\right)^\alpha~,
\end{eqnarray}
which gives the following solution for $\kappa$:
\begin{equation}
\kappa=2(\alpha-1)^2~.
\end{equation}
The brane tension, in terms of the parameters of the solution,
is therefore given by:
\begin{equation}\label{CC}
\widetilde\Lambda =2\left({\alpha-1\over \epsilon}\right)^2
\left({\epsilon\over r_B}\right)^{2\alpha}~.
\end{equation}
Note that the brane tension is always positive (assuming that the
``brane-width'' $\epsilon$ is non-zero). At first it might seem that
it can be arbitrary as we can adjust the integration constant $r_B$. However,
we also have a relation between the Gauss-Bonnet coupling $\xi$ and $r_B$:
\begin{equation}\label{GB-coupling}
 \pm\, \xi=-{\widetilde L\epsilon\over 4(\alpha -1)}    
\left({r_B\over \epsilon}\right)^\alpha~,
\end{equation}
which implies that 
\begin{equation}\label{CC-string}
 {\widetilde \Lambda}={8{\widetilde L}^2\over\xi^2}~.
\end{equation}
That is, these solutions are {\em not} ``diluting'', rather they exist
only if we fine-tune the brane tension and the Gauss-Bonnet coupling
(we discuss the reason for this in the next subsection). In the
following we will see that this specific to the case at hand, and diluting
solutions do exist in other cases.

{}Before we end this subsection, let us analyze the singularity structure
in the above solutions. Singularities can potentially occur at $r\rightarrow
\infty$ in Solution A and at $r\rightarrow 0$ in Solution B. 
Thus, the line element in the non-trivially warped part of the space-time
is given by: 
\begin{equation}
ds^2=  \left({r_A\over r}\right)^{2(\alpha-1)} \eta_{\mu\nu}dx^\mu dx^\nu
+\left({r_B\over r}\right)^{2\alpha} \delta_{ij} dx^i dx^j ~,
\end{equation}
which is singular at $r=0$ for both roots $\alpha=\alpha_\pm$. 
However, only for $\alpha=\alpha_-$ is the space truly singular, whereas for 
$\alpha=\alpha_+$ we merely have a coordinate singularity. Thus, consider 
a $2n$-derivative scalar. Such an object - let us call it $\chi^{(n)}$ -
has the following expression in terms of the extra-space radius
\begin{equation}\label{n-scalar}
\chi^{(n)}\sim {\rm e}^{-2n B} {1\over r^{2n}} \sim r^{2n(\alpha -1)}.
\end{equation}
The latter blows up for $\alpha=\alpha_-$
as $r$ approaches zero; this singularity is a naked singularity. One can indeed
consider radial null geodesics with affine parameter $\sigma$ and use that 
$G_{00}dt/d\sigma$ is constant along geodesics to obtain
\begin{equation}
{dr\over d\sigma} \sim r^{2\alpha-1}~. 
\end{equation}
For $\alpha=\alpha_-$ these geodesics terminate with finite affine parameter as
 $r$ approaches zero,
\begin{equation}
\sigma \sim r^{2(1-\alpha)}+{\rm constant}~,
\end{equation} 
Thus, we have incomplete geodesics reaching a point of 
divergent curvature. On the other hand, for $\alpha=\alpha_+$ the expression
(\ref{n-scalar}) vanishes as $r$ approaches zero, the aforementioned 
geodesics are complete ({\it i.e.}, $\sigma\rightarrow\infty$), and it is not 
difficult to see, by  doing a similar calculation, that radial time-like 
geodesics extend to infinite proper time in this limit. Therefore, 
$r=0$ is a coordinate singularity in this case. This then implies that for 
Solution B we must choose $\alpha=\alpha_+$. It then follows that the 
Gauss-Bonnet coupling $\xi$ is {\em positive} in this case.

{}Similar considerations apply to the $r\rightarrow\infty$ singularity. 
In this case we have a naked singularity for $\alpha=\alpha_+$, while for
$\alpha=\alpha_-$ we merely have a coordinate singularity. This implies that
for Solution A we must choose $\alpha=\alpha_-$. Note that in this case
the Gauss-Bonnet coupling $\xi$ is also {\em positive}.

{}Another comment we would like to make here concerns the volume of the extra
space, which is infinite. Indeed, this volume is given by
\begin{equation}
 \int_0^\infty dr~r^{d-1}~e^{{\overline d} A}~e^{dB}~.
\end{equation}
In Solution B this volume is clearly infinite. In fact, in Solution B even
the volume of the space inside of the $(d-1)$-sphere (that is, for
$0\leq r\leq \epsilon$) is infinite:
\begin{equation}
 \int_0^\epsilon dr~r^2~\left({r_A\over r}\right)^{2(\alpha_+ -1)}~
 \left({r_B\over r}\right)^{3\alpha_+}\sim\int_0^\epsilon dr~ r^{-1-{5\over
 \sqrt{3}}}~.
\end{equation}
On the other hand, in Solution A the volume of the space inside of the
$(d-1)$-sphere is finite; but the volume outside of the $(d-1)$-sphere (that 
is, for $r\geq \epsilon$) is infinite:
\begin{equation}
 \int_\epsilon^\infty dr~r^2~\left({r_A\over r}\right)^{2(\alpha_- -1)}~
 \left({r_B\over r}\right)^{3\alpha_-}\sim\int_\epsilon^\infty dr~ 
r^{-1+{5\over \sqrt{3}}}~.
\end{equation}
That  is, in both Solution A and Solution B the extra space  has
infinite volume\footnote{Note that, in Solution A, the extra space 
would appear to have finite volume had we chosen $\alpha=\alpha_+$.
However, as we saw above, in this case we would have a naked singularity at 
$r\rightarrow \infty$, and the finiteness of the volume of the extra space
would be due to truncating the space, which is geodesically incomplete, at the
naked singularity. A similar situation will arise again in the
following sections.}.
 
{}Thus, as we see, we have sensible infinite-volume 
{\em non-singular} solutions if we take the
bulk action to be given by the Gauss-Bonnet combination. However, as we
have already mentioned, these solutions exist only for the fine-tuned value
of the brane tension. If the brane tension is not fine-tuned, then we expect to
have solutions where the non-compact part of the brane is inflating.

\subsection{Membrane in 6D Bulk}
   
Let us now consider the case of a membrane (${\overline d}=3$) in 6D bulk
($d=3$). As in the string case discussed in the previous subsection 
a drastic simplification of the 
equations of motion occurs if we neglect the Einstein-Hilbert term in the bulk 
action. In this subsection we will focus on this case. 
The ($rr$), ($rr-\alpha\beta$) and ($\mu\nu$) equations of motion read:
\begin{eqnarray}
 &&A^\prime\left(2A^\prime~{\rho^\prime\over \rho}+
 3\left(\rho^\prime\over \rho\right)^2
 -{1\over r^2}\right)=0~, \label{mem1}\\
 &&2N\left(N-M\right)+(A^\prime)^2\left(S-P+N-M\right)=-{1\over 24\xi}
 {\rm e}^{3B}{\widetilde\Lambda}{\widetilde L}\delta(r-\epsilon)~,
 \label{mem2}\\
 &&2N(N+2P)+(A^\prime)^2(2P+S)+2M(2N+S)={1\over 8\xi}{\rm e}^{3B}\left(
 {\widetilde \Lambda-2\lambda}\right){\widetilde L}\delta(r-\epsilon)~.
\label{mem3}
\end{eqnarray}
Here 
\begin{equation}
 \rho\equiv r~e^B~.
\end{equation}
If we use $\rho$ instead of $r$, then the metric takes the following form:
\begin{equation}
 ds^2=\exp(2A)~\eta_{\mu\nu}~dx^\mu dx^\nu+\exp(2C)~d\rho^2+\rho^2~
 \gamma_{\alpha\beta}~dx^\alpha dx^\beta~,
\end{equation}
where, as before, $\gamma_{\alpha\beta}$ is the metric on a unit 
$(d-1)$-sphere, while $\exp(C)\equiv\exp(B)/\rho^\prime$.

{}For our purposes here it will be convenient to use $U\equiv\exp(-C)=
r~\rho^\prime/\rho$. Then from (\ref{mem1}) and (\ref{mem2}) we have (for
convenience here we also include the equation for $B$):
\begin{eqnarray}\label{mem1-fin}
 &&A^\prime={1\over 2rU}(1-3U^2)~,\\
 \label{mem2-fin}
 &&B^\prime={1\over r}(U-1)~,\\
 &&\left[r(9U^4+2U^2+1)~U^\prime+{1\over 2}
 (-33U^6+5U^4+5U^2-1)\right]{A^\prime\over U^3}=
 -{1\over 6\xi}\epsilon^3{\rm e}^{3B}{\widetilde \Lambda}{\widetilde L}
 \delta(r-\epsilon)~.
\end{eqnarray}   
The last equation can be integrated for $r\not=\epsilon$. Thus, we have
the following solution:
\begin{eqnarray}\label{solution:M-brane}
 &&{\rm ln}\left({r\over r_0}\right)=-2\sqrt{3}~{\rm arctanh}
 \left(\sqrt{3}\,U(r)\right)+{1\over v_+}{\rm arctan}\left({U(r)\over u_+}
 \right)+{1\over v_-}~
 {\rm arctanh}\left({U(r)\over u_-}\right)~,
\end{eqnarray}
where $r_0$ is an integration constant, and
\begin{eqnarray}
 &&u_\pm\equiv{\sqrt{\pm 1+2\sqrt{3}\over 11}}~,\\
 &&v_\pm\equiv{\sqrt{\pm 11+22\sqrt{3}} \over \mp 8+6\sqrt{3}}~.
\end{eqnarray}
Note that $u_+u_-=1/\sqrt{11}$, and $v_+v_-=\sqrt{11}/4$.

{}Next, note that in both the $r\rightarrow 0$ as well as the
$r\rightarrow\infty$ limits the r.h.s of (\ref{solution:M-brane}) 
is dominated by the last term, so in these limits we have
\begin{equation}
 {\rm ln}\left({r\over r_0}\right)^{v_-}\approx 
 {\rm arctanh}\left({U(r)\over u_-}\right)~,
\end{equation}
and
\begin{eqnarray}\label{limits}
 U(r)\approx \pm u_-\left[1 -2\left({r_0\over r}\right)^{\pm 2 v_-}\right]~,
\end{eqnarray}
where the ``$+$'' sign corresponds to the $r\rightarrow \infty$ limit, 
whereas the ``$-$'' sign corresponds to the $r\rightarrow 0$ limit. 
In particular, the $2n$-derivative scalar 
$\chi^{(n)}$ introduced in the previous subsection reads
\begin{eqnarray}
\chi^{(n)}\sim r^{\mp 2n u_-}\rightarrow 0~,
\end{eqnarray}
and, therefore, the solution implicitly given in (\ref{solution:M-brane})
is smooth at $r=0$ and $r\rightarrow\infty$ (since the corresponding 
singularities are coordinate singularities). In the above sign 
conventions we have the following asymptotic behavior for the warp factors,
\begin{eqnarray}
A&\sim&\pm{1-3u_-^2\over 2u_-}{\rm ln}\left(r\over r_A\right)~,\\
B&\sim&(\pm u_--1){\rm ln}\left(r\over r_B\right)~.
\end{eqnarray} 
Thus, we 
can check that null and time-like radial geodesics are complete in these 
limits. The affine parameter for null geodesics diverges as
\begin{equation}
\sigma\sim r^{\pm \left({1-u_-^2\over 2u_-}\right)}\rightarrow
\infty~,
\end{equation} 
while the proper time diverges as
$\tau\sim r^{\pm u_-}\rightarrow\infty$.

{}What about finite distance singularities? First, note that for the function
$U(r)$ in (\ref{solution:M-brane}) we have
\begin{equation}
 U^\prime={11\over 2r}~{(3U^2-1)(U^2-u_-^2)(U^2+u_+^2)\over{9U^4+2U^2+1}}~.
\end{equation}
Since $3u_-^2<1$, the l.h.s. of this equation is strictly positive for
$U$ between $-u_-$ and $+u_-$. This together with (\ref{limits}) implies that
$U(r)$ is a monotonically increasing function bounded from below by its value
$-u_-$ at $r=0$ and bounded from above by its asymptotic value $+u_-$ at 
$r\rightarrow \infty$. In particular, $U(r)$ is finite. However, $U(r)$ does
go through zero at a finite value of $r$. According
to (\ref{mem1-fin}) and (\ref{mem2-fin}) this could potentially lead to 
to a nasty finite-distance singularity in the full solution\footnote{As 
$r\rightarrow r_0$, the curvature scalar diverges as $R\sim (r-r_0)^{-2}$, and 
radial null geodesics are incomplete.}.
However, as we will now see, this does {\em not} actually take place for a 
range of values of the brane tension.

{}Thus, we need to accommodate the jump conditions 
\begin{eqnarray}\label{jump1}
 &&\left. {1\over U^3}(-1+3U^2)^2 (1+3U^2)
 \right|^{\epsilon+}_{\epsilon-}={\epsilon^3 \over \xi}~ 
 {\rm e}^{3B(\epsilon)} {\widetilde \Lambda}{\widetilde L}~,\\
 &&\left.{1\over U}(1+U^2)(-1+3U^2)\right|^{\epsilon+}_{\epsilon-}=
 {\epsilon^3\over 4\xi}~ 
 {\rm e}^{3B(\epsilon)}({\widetilde \Lambda}-2\lambda)
 {\widetilde L}~,\label{jump2}
\end{eqnarray}
which follow from (\ref{mem2}) and (\ref{mem3}). To do this, we can consider
two types of solutions:\\
$\bullet$ Solution A:
\begin{equation}
 r<\epsilon:~~~A={\rm constant}~,~~~B={\rm constant}~(U=1)~.
\end{equation}
$\bullet$ Solution B:
\begin{equation}
 r>\epsilon:~~~A={\rm constant}~,~~~B={\rm constant}~(U=1)~.
\end{equation}
We must check that we can consistently glue these solutions with the 
non-trivial solution with $U(r)$ given by (\ref{solution:M-brane}) for
$r>\epsilon$ for Solution A and for $r<\epsilon$ for Solution B.

{}Let us start with Solution B. The jump condition (\ref{jump1}) then reads:
\begin{eqnarray}\label{M-jump-1}
 \left.{1\over U^3}(-1+3U^2)^2 (1+3U^2)
 \right|_{\epsilon-}={16}-{\kappa\epsilon\over \xi}~ 
 {\rm e}^{B(\epsilon)}
{\widetilde L}~.
\end{eqnarray} 
Now consider solutions with $U(\epsilon-)<0$ (but $U(\epsilon-)>-u_-$). From 
our previous discussion it
follows that such a solution is non-singular. Assuming that ${\widetilde
\Lambda}>0$, this requires that $\xi$ be {\em positive} - indeed, the l.h.s. of
(\ref{M-jump-1}) is negative. We can then choose $B(\epsilon)$ such that 
(\ref{M-jump-1}) is satisfied (note that $U$ does not depend on $B(\epsilon)$
since it is only related to  $B^\prime$). Next, the second jump condition 
(\ref{jump2}) reads:
\begin{eqnarray}\label{M-jump-2}
 \left.{1\over U}(1+U^2)(-1+3U^2)\right|_{\epsilon-}=
 4-{(\kappa-2)\epsilon\over 4\xi}~ 
 {\rm e}^{B(\epsilon)}
 {\widetilde L}~.
\end{eqnarray}
The l.h.s. of this equation is positive. This condition then can be 
satisfied if we assume that $\kappa<2$ (note, however, that this condition is
not necessary\footnote{In fact, $\kappa>2$ values are ``phenomenologically'' 
most interesting - see the end of this section. However, since here we are 
discussing a membrane (and not a 3-brane), we will be somewhat cavalier.}). 
In fact, this solution is {\em diluting}, that is,
there is a continuous range of positive values of ${\widetilde\Lambda}$ for 
which this solution
exists. To see this, consider the case where $|U(\epsilon-)|\ll 1$. It is 
then not difficult to see that we have 
\begin{eqnarray}
 &&\kappa\approx 2~,\\
 &&{{\widetilde L}\over \xi}~\epsilon~e^{B(\epsilon)}\approx{1\over 
 2|U^3(\epsilon-)|}~.
\end{eqnarray}
This implies that
\begin{equation}
 {\widetilde\Lambda}=\kappa\lambda\approx U^6(\epsilon-)~{8{\widetilde L}^2
 \over\xi^2}~.
\end{equation}
Note the difference between this result and the corresponding result 
(\ref{CC-string}) for the
string discussed in the previous subsection. In the latter case the
brane tension had to be fine-tuned. In the case of the membrane we actually
have {\em diluting} - the solution exists for a range of the brane
tension ${\widetilde\Lambda}$. Indeed, $U(\epsilon-)$ depends on the
integration constant $r_0$, which is the source of this diluting 
property. Thus, in the case where $|U(\epsilon-)|\ll 1$ we have
\begin{equation}
 U(\epsilon-)\approx -{1\over 2}~\ln\left({r_0\over\epsilon}\right)~,
\end{equation}
so that the brane tension ${\widetilde \Lambda}$ depends on $r_0$. Note that
in the case of the string the function $U(r)$ defined as above is actually a
{\em numerical constant} away from the brane. This is precisely why we did
not find a diluting solution in the case of the string. On the other hand, for
a membrane as well as higher dimensional branes this function is non-trivial
and depends on an additional integration constant, hence the diluting property
of such solutions. In turn, the reason why this is possible, say, in the 
membrane case is that equations of motion are more non-linear than those
in the string case.

{}Let us now consider Solution A. In this case the jump conditions read:
\begin{eqnarray}\label{M-jump-1-r}
 &&\left.{1\over U^3}(-1+3U^2)^2 (1+3U^2)
 \right|_{\epsilon+}={16}+{\kappa\epsilon\over \xi}~ 
 {\rm e}^{B(\epsilon)}
 {\widetilde L}~,\\\label{M-jump-2-r}
 &&\left.{1\over U}(1+U^2)(-1+3U^2)\right|_{\epsilon+}=
 4+{(\kappa-2)\epsilon\over 4\xi}~
 {\rm e}^{B(\epsilon)}
 {\widetilde L}~.
\end{eqnarray}
If we take $U(\epsilon+)>0$ (but $U(\epsilon+)<u_-$), then we have a 
smooth solution. In fact, this solution is also {\em diluting}. This can be
seen by noting that choosing {\em positive} $\xi$ and $\kappa<2$ ensures that
the jump conditions can be satisfied. 

{}The last comment we would like to make here concerns the volume of the extra
space. It is evident that this volume is infinite in Solution B. For solution A
we have the following expression for this volume:
\begin{equation}
 \int_0^\infty dr~r^2~e^{3A}~e^{3B}~.
\end{equation} 
Asymptotically at $r\rightarrow \infty$ we have
\begin{eqnarray}
 &&A(r)\sim {(1-3u_-^2)\over 2u_-}\ln\left({r\over r_A}\right)~,\\
 &&B(r)\sim (u_- -1)\ln\left({r\over r_B}\right)~.
\end{eqnarray}
It is then not difficult to see that the volume of the extra space in Solution
A is infinite as well. Numerical results (obtained using Maple 6) 
for $A(r)$ and $B(r)$ for Solution A
are presented in Fig.1 and Fig.2, respectively.

{}Thus, as we see, we have sensible infinite-volume {\em non-singular}
solutions in the case of the membrane as well. Unlike in the string case,
however, these solutions have the {\em diluting} property - they exist for
a continuous range of the brane tension values.

\subsection{3-Brane in 7D Bulk}
   
{}In this subsection we consider the case of a 3-brane (${\overline d}=4$) 
in 7D bulk ($d=3$). As in the previous subsections we will neglect the 
Einstein-Hilbert term in the bulk action. Let
\begin{equation}
 U\equiv rB^\prime+1~.
\end{equation}
Then the $(rr)$ equation of motion reads:
\begin{equation}
\label{3brr}
 (A^\prime)^2\left[(A^\prime)^2+8A^\prime~{U\over r}
 +2~{3U^2-1\over r^2}\right]=0~,
\end{equation}
which implies that either $A^\prime=0$ or 
\begin{equation}
\label{3bA}
 rA^\prime=-4U\pm \sqrt{10U^2+2}\equiv -4U\pm T~.
\end{equation}    
The ($rr-\alpha\beta$) and ($\mu\nu$) equations of motion 
read:
\begin{eqnarray}\label{alphabeta}
 &&2\left[M-N\right]\left[(A^\prime)^2+N\right]+(A^\prime)^2\left[P-S\right]=
 {1\over 48\xi}~
 e^{3B}{\widetilde\Lambda}{\widetilde L}\delta(r-\epsilon)~,\\
 &&2\left[(A^\prime)^2+N\right]\left[P+N\right]+
 M\left[4N+S+(A^\prime)^2\right]+
 (A^\prime)^2 S={1\over 24\xi}~
 e^{3B}\left({\widetilde\Lambda}-2\lambda\right)
 {\widetilde L}\delta(r-\epsilon)~.
\end{eqnarray}
Note that
\begin{eqnarray}
 &&P={1\over r}~U^\prime~,\\
 &&S={1\over r^2}~\left(U^2-1\right)~,\\
 &&N={1\over r}~A^\prime U={1\over r^2}~\left[\eta TU-4U^2\right]~,\\
 &&M={U^\prime\over rT}~\left[10\eta U-4T\right]+{1\over r^2}~
 \left[30U^2-9\eta TU+2\right]~,
\end{eqnarray}
where $\eta=\pm 1$. Using these expressions we can rewrite (\ref{alphabeta})
as follows: 
\begin{eqnarray}\label{3bU}
 &&(\eta T-4U)\left\{rU^\prime\left[40U-13\eta T+{12\eta\over T}\right]-
 \left[400 U^3-127\eta T U^2+56 U-5\eta T\right]\right\}
 =\nonumber\\
 &&{\kappa\epsilon^2\over 48\xi}~e^B {\widetilde L} \delta(r-\epsilon),
\end{eqnarray}
Thus, away from the brane the function $U$ satisfies the following first order
differential equation (assuming a non-trivial solution, that is $A\not=
{\rm const.}$):
\begin{equation}
 rU^\prime={400 U^3-127\eta T U^2+56 U-5\eta T\over
 40U-13\eta T+{12\eta/ T}}~.
\end{equation}
It is not difficult to show that for $\eta=+1$ the denominator is always 
negative, while for $\eta=-1$ it is always positive. As to the numerator, it
is positive for $U<u_1$ and $U>u_2$, it is negative for $u_1<U<u_2$, and it
vanishes for $U=u_1$ and $U=u_2$, where
\begin{eqnarray}
 &&\eta=-1:~~~u_1=u^-_1\approx -.58~,~~~u_2=u^-_2\approx -.43~,\\
 &&\eta=+1:~~~u_1=u^+_1\approx +.43~,~~~u_2=u^+_2\approx +.58~.
\end{eqnarray}
This implies that we have the following solutions for $U(r)$, all of which are
{\em monotonically increasing}:
\begin{eqnarray}
 &&\eta=-1:~~~U(r\rightarrow r_0+)\rightarrow -\infty~,~~~U(r\rightarrow\infty)
 \rightarrow u^-_1~,\\
 &&\eta=-1:~~~U(r\rightarrow 0)\rightarrow u^-_2~,~~~U(r\rightarrow r_0-)
 \rightarrow +\infty~,\\
 &&\eta=+1:~~~U(r\rightarrow 0)\rightarrow u^+_1~,~~~U(r\rightarrow\infty)
 \rightarrow u^+_2~.
\end{eqnarray}
Here $r_0$ in each of the $\eta=-1$ cases is an integration constant (see 
below). Let us discuss these solutions in more detail.

\newpage
\begin{center}
 {\em The $\eta=+1$ Solutions} 
\end{center}

{}In this solution the $r$
dependence of $U$ is given by
\begin{equation}
 U(r)=f(\ln(r/r_0))~,
\end{equation}
where $r_0$ is an integration constant, while the function $f(x)$ is the 
solution of the differential equation
\begin{equation}
 {df\over dx}={400 f^3-127 W f^2+56 f-5 W\over
 40f-13 W+{12/ W}}~,
\end{equation}
where $W\equiv\sqrt{10f^2+2}$, and the boundary conditions are chosen as
$f(x\rightarrow-\infty)=u^+_1$ and $f(x\rightarrow+\infty)=u^+_2$. Note that
$f(x)$ is a monotonically increasing function.

{}Next, to find a full solution, we must impose the jump conditions. These are
given by:
\begin{eqnarray}
 &&\left.\left[{46\over 15}~ T^3-{290\over 3}~U^3-14 U-{24\over 5}~ T
 \right]\right|^{\epsilon+}_{\epsilon-}={\kappa\epsilon\over 48\xi}~
 e^{B(\epsilon)}{\widetilde L}~,\\
 &&\left.\left[{13\over 30}~ T^3-{40\over 3}~U^3-{6\over 5}~ T
 \right]\right|^{\epsilon+}_{\epsilon-}={(\kappa-2)\epsilon\over 24\xi}~
 e^{B(\epsilon)}{\widetilde L}~.
\end{eqnarray}
Let us consider solutions where $A$ and $B$ are constant for $r<\epsilon$
(in this region we then have $U=1$), and have non-trivial profiles for
$r>\epsilon$ (in this region $U$ is non-trivial). The jump conditions then
read:
\begin{eqnarray}
 &&\left.\left[{46\over 15}~ T^3-{290\over 3}~U^3-14 U-{24\over 5}~ T
 \right]\right|_{\epsilon+}=64\sqrt{3}-{332\over 3}+
 {\kappa\epsilon\over 48\xi}~
 e^{B(\epsilon)}{\widetilde L}~,\\
 &&\left.\left[{13\over 30}~ T^3-{40\over 3}~U^3-{6\over 5}~ T
 \right]\right|_{\epsilon+}=8\sqrt{3}-{40\over 3}+
 {(\kappa-2)\epsilon\over 24\xi}~
 e^{B(\epsilon)}{\widetilde L}~.
\end{eqnarray}
Let us rewrite these conditions as follows:
\begin{eqnarray}
  &&\left.\left[{46\over 15}~ T^3-{290\over 3}~U^3-14 U-{24\over 5}~ T
 \right]\right|_{\epsilon+}=64\sqrt{3}-{332\over 3}+
 {\kappa\epsilon\over 48\xi}~
 e^{B(\epsilon)}{\widetilde L}~,\\
  &&\left.\left[{57\over 20}~ T^3-90~U^3-14 U-{21\over 5}~ T
 \right]\right|_{\epsilon+}=60\sqrt{3}-104+
 {\epsilon\over 24\xi}~
 e^{B(\epsilon)}{\widetilde L}~.
\end{eqnarray}
Then one can check numerically that we do not have a solution
with positive $\kappa$ (that is, for positive ${\widetilde \Lambda}$). 
Thus, $\kappa$ ranges between about $-1.60$ and $-4.80$ depending on the
value of $U(\epsilon+)$ (which must be between $u^+_1$ and $u^+_2$), 
while $\epsilon \exp[B(\epsilon)]$ ranges between 
about 4.45 and 1.85 times $\xi/{\widetilde L}$, so that ${\widetilde\Lambda}$
ranges between about $-.080$ and $-1.41$ times ${\widetilde L}^2/\xi^2$. 
That is, this solution is {\em diluting}. Note
that in this solution $\xi$ must be {\em positive}. Also, 
the volume of the extra space in this solution is 
infinite. Thus,
the volume of the extra space is given by:
\begin{equation}
 \int_0^\infty dr~r^2~e^{4A}~e^{3B}~.
\end{equation}  
On the other hand, for large $r$ we have
\begin{eqnarray}
 &&A(r)\sim (\sqrt{10 (u^+_2)^2+2}-4u^+_2)\ln\left({r\over r_A}\right)~,\\
 &&B(r)\sim (u^+_2-1)\ln\left({r\over r_B}\right)~.
\end{eqnarray}
Then it is not difficult to show that the volume of the extra space is
infinite, albeit the
brane tension must be negative.

{}Alternatively, we can consider solutions where $A$ and $B$ are constant for
$r>\epsilon$. Note that in these solutions the volume of the extra space is
automatically infinite. The matching conditions now read:
\begin{eqnarray}
 &&\left.\left[{46\over 15}~ T^3-{290\over 3}~U^3-14 U-{24\over 5}~ T
 \right]\right|_{\epsilon-}=64\sqrt{3}-{332\over 3}-
 {\kappa\epsilon\over 48\xi}~
 e^{B(\epsilon)}{\widetilde L}~,\\
  &&\left.\left[{57\over 20}~ T^3-90~U^3-14 U-{21\over 5}~ T
 \right]\right|_{\epsilon-}=60\sqrt{3}-104-
 {\epsilon\over 24\xi}~
 e^{B(\epsilon)}{\widetilde L}~.
\end{eqnarray}
It is not difficult to show that in these solutions, which are {\em diluting}
with negative brane tension, $\xi$ must be {\em negative}.  

\begin{center}
 {\em The $\eta=-1$ Solutions}
\end{center}

{}Let us now discuss the $\eta=-1$ solutions. If we choose the solution such
that $U(r\rightarrow \infty)\rightarrow u^-_1$, then the $r$
dependence of $U$ is given by
\begin{equation}
 U(r)=f(\ln(r/r_0))~,
\end{equation}
where $r_0$ is an integration constant, while the function $f(x)$ is the 
solution of the differential equation
\begin{equation}
 {df\over dx}={400 f^3+127 W f^2+56 f+5 W\over
 40f+13 W+{12/ W}}~,
\end{equation}
where $W\equiv\sqrt{10f^2+2}$, and the boundary conditions are chosen as
$f(x\rightarrow 0+)\rightarrow -\infty$ and 
$f(x\rightarrow+\infty)=u^-_1$. Note that
$f(x)$ is a monotonically increasing function.
To ensure the absence of
a singularity at $r=r_0$ (see below)
we can consider the solution where $A$ and $B$ are
constant for $r<\epsilon$, where $\epsilon>r_0$. 
However, it is not difficult to show that the
volume of the extra space in this solution is finite. 

{}A solution with
infinite-volume extra space can be obtained if we take $A$ and $B$ to be 
constant for $r>\epsilon$. In this solution the $r$
dependence of $U$ is given by
\begin{equation}
 U(r)=f(\ln(r/r_0))~,
\end{equation}
where $r_0$ is an integration constant, while the function $f(x)$ is the 
solution of the differential equation
\begin{equation}
 {df\over dx}={400 f^3+127 W f^2+56 f+5 W\over
 40f+13 W+{12/ W}}~,
\end{equation}
where $W\equiv\sqrt{10f^2+2}$, and the boundary conditions are chosen as
$f(x\rightarrow-\infty)=u^-_2$ and $f(x\rightarrow 0-)=+\infty$. Note that
$f(x)$ is a monotonically increasing function. To ensure the absence of a 
singularity at $r=r_0$ (see below) we must choose $\epsilon<r_0$.
The matching conditions then read:
\begin{eqnarray}
  &&\left.\left[-{46\over 15}~ T^3-{290\over 3}~U^3-14 U+{24\over 5}~ T
 \right]\right|_{\epsilon-}=-64\sqrt{3}-{332\over 3}-
 {\kappa\epsilon\over 48\xi}~
 e^{B(\epsilon)}{\widetilde L}~,\\
  &&\left.\left[-{57\over 20}~ T^3-90~U^3-14 U+{21\over 5}~ T
 \right]\right|_{\epsilon-}=-60\sqrt{3}-104-
 {\epsilon\over 24\xi}~
 e^{B(\epsilon)}{\widetilde L}~.
\end{eqnarray}
In this solution $\xi$ must be {\em negative} for $U(\epsilon-)<1$, while for
$U(\epsilon-)>1$ it must be {\em positive} (note that $U(\epsilon-)$ in this
solution can take values larger than $u^-_2$). 
That is, we have two different
{\em branches} here.
The parameter $\kappa$ is not
very sensitive to the value of $U(\epsilon-)$, in particular, it
smoothly changes from the $\xi<0$ branch to the $\xi>0$ branch - 
it ranges between about
2.13 if $U(\epsilon-)$ is close to $u^-_2$ and 2.15 if $U(\epsilon-)\gg 1$. 
As to ${\widetilde \Lambda}$, if $U(\epsilon-)<1$, 
it grows from about $8.57\times 10^{-8}~
({\widetilde L}^2/\xi^2)$ if $U(\epsilon-)$ is close to $u^-_2$ up to infinity
as $U(\epsilon-)$ approaches 1. On the other hand, if $U(\epsilon-)>1$,
${\widetilde \Lambda}$ decreases from infinity, as $U(\epsilon-)$ moves away
from 1, down to zero. Thus, we have a {\em diluting} solution with 
{\em positive} brane tension.

\begin{center}
 {\em Singularity Structure}
\end{center}

{}Before we end this subsection, let us analyze the singularity structure of
the above solutions. In particular, if $U\rightarrow U_*$ in any given
solution for $r\rightarrow 0$ or $r\rightarrow\infty$, then 
the $2n$-derivative scalar introduced in the previous subsections behaves as
\begin{equation}
 \chi^{(n)}\sim e^{-2nB}{1\over r^{2n}}\sim r^{-2nU_*}~,
\end{equation}
while the affine parameter $\sigma$ and proper time $\tau$, for null and 
time-like geodesics, respectively, have the following asymptotic 
expressions for the $\eta=\pm 1$ solutions:
\begin{eqnarray}
\sigma &\sim&r^{-3U_*\pm\sqrt{10U_*^2+2}}+{\rm constant}~,\\
\tau&\sim&r^{U_*}+{\rm constant}~.
\end{eqnarray}   
In the $\eta=+1$ solution $U_*=u^+_1$ for $r\rightarrow 0$ and $U_*=u^+_2$
for $r\rightarrow\infty$. We therefore would have a coordinate singularity at
$r\rightarrow \infty$ as $\chi^{(n)}\rightarrow 0$ and $\lambda$ and $\tau$ 
diverge, but at $r\rightarrow 0$ we would have a true 
singularity as we have diverging curvature and incomplete geodesics. This 
implies that for $\eta=+1$ we must choose the
solution where $A$ and $B$ are
constant for $r<\epsilon$. Recall that in this solution $\xi$ must be {\em
positive} (as the solution with negative $\xi$ has a true singularity at 
$r=0$).

{}In the $\eta=-1$ case the solution with $U(r\rightarrow\infty)\rightarrow
u^-_1$ has a true singularity at $r\rightarrow \infty$. This explains why
in this solution we found that the volume of the extra space is finite.
This appears to be an artifact of cutting off the extra space at the
singularity. At any rate, since this solution is truly singular, we will no
longer focus on it.

{}On the other hand, the $\eta=-1$ solution with $U(r\rightarrow 0)
\rightarrow u^-_2$ only has a coordinate singularity at $r=0$, so this 
solution is perfectly acceptable. In this solution, which is {\em diluting} 
with {\em positive} brane tension, the volume of the extra 
space is infinite. Numerical results for this solution are given in Fig.3 and
Fig.4.

{}Finally, let us show that at $r\rightarrow r_0\pm$ in the
the $\eta=-1$ solutions where $U(r\rightarrow r_0+)\rightarrow-\infty$ and
$U(r\rightarrow r_0-)\rightarrow+\infty$ we would indeed have true
singularities (that is, if we do not replace them by constant $A$ and $B$ 
solutions). In the former case as $r\rightarrow r_0+$ we have
\begin{equation}
 rU^\prime\sim\alpha U^2~,
\end{equation}
where
\begin{equation}
 \alpha\equiv {{400-127\sqrt{10}}\over{40-13\sqrt{10}}}\approx 1.45~.
\end{equation}
It then follows that
\begin{eqnarray}
 &&U\sim -{1\over\alpha\ln(r/r_0)}~,\\
 &&B\sim -{1\over\alpha}~\ln\left(\ln(r/r_0)\right)~,
\end{eqnarray}
so that as $r\rightarrow r_0+$ the curvature scalar diverges as 
\begin{equation}
 R\sim\left(\ln(r/r_0)\right)^{{2\over\alpha}(1-\alpha)}~,
\end{equation}
and radial null geodesics terminate with finite affine parameter at $r=r_0$.
Similarly, in the latter case as $r\rightarrow r_0-$ we have
\begin{equation}
 rU^\prime\sim\beta U^2~,
\end{equation}
where
\begin{equation}
 \beta\equiv {{400+127\sqrt{10}}\over{40+13\sqrt{10}}}\approx 9.88~.
\end{equation}
It then follows that
\begin{eqnarray}
 &&U\sim {1\over\beta\ln(r_0/r)}~,\\
 &&B\sim -{1\over\beta}~\ln\left(\ln(r_0/r)\right)~,\label{sing}
\end{eqnarray}
so that as $r\rightarrow r_0-$ the curvature scalar diverges as 
\begin{equation}
 R\sim \left(\ln(r_0/r)\right)^{{2\over\beta}(1-\beta)}~.
\end{equation}
and radial null geodesics are incomplete reaching $r=r_0$ within finite 
affine parameter. 

{}Thus, as we see, we have a sensible 
{\em diluting} non-singular solution, namely,
the $\eta=-1$ solution with infinite volume and positive brane tension.
In the next subsection we will discuss this solution in the context of the
cosmological constant problem.

\subsection{The Brane Tension}

{}The last solution we discussed in the previous subsection, namely,
that corresponding to the $\xi>0$ branch, is particularly interesting. 
This solution is {\em diluting} with positive brane tension, so we would like
to discuss it in a bit more detail. In fact, here we would like to give a 
general discussion of certain points relevant to the cosmological constant
problem. 

{}Thus, we have been referring to ${\widetilde\Lambda}$ (or, more precisely,
${\widetilde T}\equiv 
{\widetilde M}_P^{D-3}{\widetilde\Lambda}$) as the brane tension. This 
quantity is indeed the tension of the brane $\Sigma$ whose world-volume has
the geometry of ${\bf R}^{{\overline d}-1,1}\times {\bf S}^{d-1}_\epsilon$. 
Note, however, that a bulk observer at $r>\epsilon$ does not see this brane
tension - to such an observer the brane appears to be tensionless. Indeed,
the warp factors are constant in the aforementioned solution at $r>\epsilon$.
The non-vanishing (in fact, positive) tension of the brane 
$\Sigma$ does not curve the space outside of the sphere 
${\bf S}^{d-1}_\epsilon$. Instead, it curves the space {\em inside} the sphere 
${\bf S}^{d-1}_\epsilon$, that is, at $r<\epsilon$. And this happens without
producing any singularity at $r<\epsilon$, and with the non-compact part of the
world-volume of the brane remaining flat.

{}Here it is important to note that the effective tension of the 
{\em fat} $({\overline d}-1)$-brane whose world-volume is 
${\bf R}^{{\overline d}-1,1}$ (this $({\overline d}-1)$-brane 
is fat as it is extended in the
extra $(d-1)$ angular dimensions) is also positive. 
It is not difficult to see that this brane tension is given by
\begin{equation}
 {\widehat T}=({\widetilde\Lambda}-(d-1)\lambda)v_{d-1}
 {\widetilde M}_P^{D-3}~,
\end{equation}
where $v_{d-1}=\epsilon^{d-1} e^{(d-1)B(\epsilon)} 
{\overline v}_{d-1}$ is the volume of the
sphere ${\bf S}^{d-1}_\epsilon$, ${\overline v}_{d-1}$ is the volume of the
{\em unit} $(d-1)$-sphere, and we have taken into account that the radius of
the sphere ${\bf S}^{d-1}_\epsilon$ is not $\epsilon$ but rather 
\begin{equation}
 R\equiv\epsilon e^{B(\epsilon)}~. 
\end{equation}
We therefore have:
\begin{equation}
 {\widehat T}=(d-2)(\kappa-(d-1)){\overline v}_{d-1} R^{d-3}
 {\widetilde M}_P^{D-3}~.
\end{equation}
Thus, for $\kappa>2$ (which is part of the parameter space for the 
aforementioned solution) this effective fat brane tension is {\em positive}.

{}Next, the ${\overline d}$-dimensional Planck scale ${\widehat
M}_P$ is given by
\begin{equation}
 {\widehat M}_P^{{\overline d}-2}=v_{d-1}
 {\widetilde M}_P^{D-3}={\overline v}_{d-1} R^{d-1}
 {\widetilde M}_P^{D-3}~.
\end{equation}
In particular, if we consider a 3-brane in 7D bulk, we have:
\begin{eqnarray}
 &&{\widehat\Lambda}=4\pi(\kappa-2){\widetilde M}_P^4~,\\
 &&{\widehat M}_P^2=4\pi R^2 {\widetilde M}_P^4~,
\end{eqnarray}
where ${\widehat \Lambda}$ is the 3-brane tension, ${\widehat M}_P$ is the
4-dimensional Planck scale, ${\widetilde M}_P$ is the 6-dimensional Planck
scale, and $R$ is the radius of the extra 2-sphere (recall that we actually
have a 5-brane with two dimensions curled up into a 2-sphere of radius $R$,
while the radial direction transverse to this 5-brane is non-compact and has 
infinite-volume).

{}{\em A priori} we can reproduce the 4-dimensional Planck scale 
${\widehat M}_P\sim 10^{18}~{\rm GeV}$ by choosing $R$ between 
$R\sim {\rm millimeter}$ and $R\sim 1/{\widehat M}_P$. The 6-dimensional Planck
scale then ranges between\footnote{Note a similarity with \cite{ADD}.}
${\widetilde M}_P\sim {\rm TeV}$ and ${\widetilde M}_P\sim {\widehat M}_P$.
On the other hand, it is reasonable to assume that the analyses of 
\cite{DGHS} should give the same ``see-saw'' modification of gravity in the
present case, in particular, once we take into account higher curvature
terms on the {\em brane}, 
to obtain 4-dimensional gravity up to the distance scales 
of order of the present Hubble size, we must assume that the ``fundamental''
7-dimensional Planck scale\footnote{Actually, we obtained the above solution
by neglecting the Einstein-Hilbert term in the bulk. Here we will simply 
assume that the aforementioned conclusion holds, and return to this issue
when we study the more realistic model with both Einstein-Hilbert and 
Gauss-Bonnet terms in the bulk.\label{f6}}
\begin{equation}
 M_P\sim ({\rm millimeter})^{-1}~.
\end{equation}
Let us see what range of values we can expect for the 5-brane tension
${\widetilde T}$.

{}Thus, the 5-brane tension 
\begin{equation}
 {\widetilde T}=
 {\widetilde \Lambda}{\widetilde M}_P^4=\kappa R^{-2}{\widetilde M}_P^4
 \sim R^{-4} {\widehat M}_P^2~.
\end{equation}
If we assume that the Standard Model fields come from a 6-dimensional 5-brane
theory compactified on the 2-sphere\footnote{Note that {\em a priori} we could
attempt to have the Standard Model fields living on a 3-brane inside of the
5-brane (or at an orbifold fixed locus). This, however, would generically
spoil the diluting property of the solution. Indeed, the aforementioned 
3-brane inside of the 5-brane would have its own brane tension associated with
it, which would generically 
have to be fine tuned. Another point worth mentioning here is
that compactification on a 2-sphere of radius $R\ll ({\rm TeV})^{-1}$ need not
in general result in supersymmetry breaking at a scale of order $1/R$. Thus,
one could imagine embedding our scenario in a higher dimensional theory (with
perhaps some extra dimensions compactified) where the 2-sphere is fibered
over a 4-dimensional base (such that the resulting space is supersymmetric;.
in such 11-dimensional setup we might need, say, 
an additional ${\bf Z}_2$ orbifold
to obtain a chiral 4-dimensional theory). Alternatively, the 5-brane could be
a brane wrapped on a ${\bf P}^1$ corresponding to an orbifold blow-up.}, 
then we might need to require that 
$R{\ \lower-1.2pt\vbox{\hbox{\rlap{$<$}\lower5pt\vbox{\hbox{$\sim$}}}}\ }
({\rm TeV})^{-1}$. Then ${\widetilde M}_P$ ranges between $10^7-10^8~{\rm TeV}$
and ${\widehat M}_P$, while the 5-brane tension ranges between $(10^{5}~{\rm
TeV})^6$ and ${\widehat M}_P^6$. Note that {\em a priori} this is not in
conflict with having the supersymmetry breaking scale in the TeV range.

{}In principle, the above scenario {\em a priori} does not seem to be
inconsistent modulo the fact that we still need to explain why the 
6-dimensional Plank scale ${\widetilde M}_P$ is many orders of magnitude
(30 in the extreme case where $R^{-1}\sim{\widetilde M}_P\sim {\widehat M}_P$) 
higher than the seven-dimensional Planck scale $M_P$ (see the last section
for some speculations on this point). Note, however, that the same 
issue is present in any theory with infinite-volume extra dimensions. The
diluting property of our non-singular solution could be suggestive as far as
addressing the cosmological constant problem is concerned. In particular, we
do have a solution where the non-compact 4-dimensional part of the brane
is flat, yet its tension can take positive values in a continuous range. In 
particular, no fine tuning appears to be required in our solution. 

{}Before we end this section, let us see what kind of values of $U(\epsilon-)$
we would need to have in order to obtain a solution satisfying the above
phenomenological considerations. First, we will assume that the Gauss-Bonnet
parameter $\xi\sim M_P^{-2}$ (its ``natural'' value - see footnote~\ref{f6}).
Then from the matching conditions we would obtain ($U(\epsilon-)\gg 1$):
\begin{eqnarray}
 &&\kappa R {\widetilde L}/\xi\approx 9295~U^3(\epsilon-)~,\\
 &&R {\widetilde L}/\xi\approx 4323~U^3(\epsilon-)~,
\end{eqnarray}
which gives $\kappa\approx 2.15$, and
\begin{equation}
 U(\epsilon-)\sim 10^{29}~.
\end{equation}
Here for definiteness we have assumed the extreme case $R^{-1}\sim 
{\widetilde M}_P\sim {\widehat M}_P$, where
${\widetilde L}={\widetilde M}_P^4/
M_P^5\sim 10^{120}~{\rm mm}$. This implies that $\epsilon$ is very close to
the would-be singularity $r_0$:
\begin{equation}
 {r_0\over\epsilon}-1\sim 10^{-30}~,
\end{equation}
Note, however, that the warp factor
\begin{equation}
 B(\epsilon)\sim 7~,
\end{equation}
which is due to the double logarithmic suppression in (\ref{sing}).

\section{Einstein-Hilbert and Gauss-Bonnet Terms in the Bulk}

{}In this section we discuss a more realistic setup where we have both
the Einstein-Hilbert as well as Gauss-Bonnet terms in the bulk. This 
substantially increases the complexity of the equations of motions we need
to analyze, so we warm up with the example of a string in 5D bulk, and then
turn to the most interesting case of a 3-brane in 7D bulk.

\subsection{String in 5D Bulk}

{}In this subsection we study the case of a string in five-dimensional bulk 
in the presence of both the Einstein-Hilbert and 
Gauss-Bonnet terms in the bulk action. The $(rr)$, 
$(rr)-(\alpha\beta)$ and $(\mu\nu)$ equations read, respectively:

\begin{eqnarray} 
 &&(A^\prime)^2 +4N +S -4\xi{r^2\over \rho^2}\left(2N^2+{A^\prime}^2 S
 \right)=0~,
\label{AC1} \\
 &&2\left(N-M\right)+S-P-4\xi{r^2\over\rho^2}\left[2N\left(N-M\right)+
 {A^\prime}^2
 \left(S-P\right)\right]={1\over 2}  {\rm e}^{B} {\widetilde\Lambda}
 {\widetilde L} \delta(r-\epsilon)~,\label{AC2}\\
 &&M+2N+2P+S-4\xi{r^2\over \rho^2}\left(MS+2NP\right)=-{1\over 2}{\rm e}^{B}
 \left({\widetilde\Lambda}-2\lambda\right){\widetilde L}\delta(r-\epsilon)~.
 \label{AC3}
\end{eqnarray}
A trivial solution for $r\not =\epsilon$ is given by 
$A^\prime=B^\prime=0$. To 
study non-trivial solutions of this system, we define $V\equiv rA^\prime$ and 
$z\equiv\rho^2/\xi$ (so that $2U=rz^\prime/z$, where we are using the
definitions for $U$ and $\rho$ from the previous section), 
in terms of which (\ref{AC1}) becomes:

\begin{equation}\label{AD1}
zV^2+zU^2+4zVU-12V^2U^2+4V^2-z=0~.
\end{equation} 
Thus, non-trivial solutions satisfy

\begin{eqnarray}
V&=&{-2zU+\zeta\sqrt{3z^2U^2+12zU^4-16zU^2+z^2+4z}\over z-12U^2+4}~,	\\
N&=&{VU\over r^2}~,	\\
M&=&{1\over r^2}\left(rV^\prime+V^2-VU\right),
\end{eqnarray}
where $\zeta=\pm 1$, and (\ref{AC2}) reads
\begin{equation}\label{AD2}
2(z-4UV)(2UV-V^2-rV^\prime)+(z-4V^2)(U^2-1-rU^\prime)=
{1\over 2}z\epsilon^2{\rm e}^B\t\Lambda\t L\delta(r-\epsilon)~.\end{equation}
Away from the brane this gives a first order differential equation for $U$ as 
a function of $r$, namely:

\begin{equation}\label{dudr} 
rU^\prime={f_1\over f_2 }~,
\end{equation}
or treating $U$ as function of $z$ we have:

\begin{equation}\label{dudz} 
2zU{dU\over dz}={f_1\over f_2 }~,
\end{equation}
where we have defined 
\begin{eqnarray}
 f_1&\equiv&
 48zU^2V-240zV^2U^3+4z^2V^2U+34z^2U^2V+24zU^2V^3+32V^3-8zV+\nonumber\\ 
 &&48zUV^2+8z^2U^3-8z^2U-2z^2V-256U^2V^3+480V^3U^4-8zV^3-4z^2V^3-\nonumber\\
 &&192U^3V^4+64UV^4-40zU^4V+16zUV^4~,\\
 f_2&\equiv&64zV^2U-6z^2V-96U^2V^3-8zU^2V+8zV-32V^3-8V^3z~,
\end{eqnarray}
for later convenience.

{}Note that solutions 
must satisfy the following jump conditions implied by (\ref{AD2})
and the ($\mu\nu$) equation of motion:
 
\begin{eqnarray}
\left.\left[-z(2V+U)+4V^2U\right]\right|^{\epsilon+}_{\epsilon-}&=&
{R\kappa \t L\over 2\xi}~,\\
\left.\left[4V(U^2-1)-z(2U+V)\right]\right|^{\epsilon+}_{\epsilon-}&=&
\left(\kappa -2\right){R\t L\over 2\xi}~.
\end{eqnarray}
Let us rewrite these conditions as follows: 
\begin{eqnarray}
\left.\left[4VU(V-U)+4V+z(U-V)\right]\right|^{\epsilon+}_{\epsilon-}&=&
{R \t L\over \xi}~,\label{MC1}\\
\left.\left[8V^2U-2z(U+2V)\right]\right|^{\epsilon+}_{\epsilon-}&=&
{R\t L\over \xi}\kappa~,\label{MC2}
\end{eqnarray}
so that (\ref{MC1}) only involves the coupling $\xi$, whereas the ratio
of (\ref{MC2}) and (\ref{MC1}) only depends upon $\kappa$.
We will consider two types of solutions: those that have constant 
warp factors $A$ and $B$ for 
$r<\epsilon$, and non-trivial $A$ and $B$ for 
$r>\epsilon$ - the {\it exterior} solutions; and those
that have non-trivial $A$ and $B$ for $r<\epsilon$, and constant $A$ and $B$ 
for $r>\epsilon$ - {\it interior} solutions.
Through the definition of the functions $U$ and $V$ we can rewrite (\ref{MC1}) 
as follows:

\begin{equation}
\left.\left[4VU(V-U)+4V+z(U-V-1)\right]\right|_{\epsilon\pm}=\pm
{R \t L\over \xi}~,\label{MC3}
\end{equation}  
with the plus sign corresponding to exterior solutions and the minus 
sign corresponding to interior ones. 
Fig.5 shows the regions in the $U-z$ plane where the function 
$4VU(V-U)+4V+z(U-V-1)$ is positive 
(suitable for matching exterior solutions with $\xi>0$ or interior solutions 
with $\xi<0$) and negative (suitable for the other two possibilities) for the
$\zeta=+1$ case.
We will now study the solutions with positive Gauss-Bonnet coupling $\xi$ ({\it
 i.e.}, $z>0$) since this was the case that allowed non-singular solutions in 
the absence of the Einstein-Hilbert term in the bulk action. 

\begin{center}
{\it The $\zeta=+1$ Solutions}
\end{center}

{}In order to have exterior solutions we must have either $U(\epsilon+)>1$  or 
$U(\epsilon+)$ less than the negative zero of $4VU(V-U)+4V+z(U-V-1)$ shown in 
Fig.5. In the former case $\kappa$ is strictly negative, while $\kappa$ is 
positive in the latter case. 
In both allowed regions $-1<f_1/(2zUf_2)<0$. This means that the 
{\it negative} $\kappa$ solutions flow to lower values of $U$ while $z$ 
increases as 
$r\rightarrow\infty$. A numerical study (using Maple 6) shows that as $r$ 
increases, the solutions approach the region $z\gg 1$, 
$U{\ \lower-1.2pt\vbox{\hbox{\rlap{$>$}\lower5pt\vbox{\hbox{$\sim$}}}}\ }1$.
In this region (\ref{dudz}) becomes (to the leading order in $1/z$),
\begin{equation}
-{2z\over U-1}{dU\over dz}\sim 1~,
\end{equation}
which is integrated to give
\begin{equation}\label{AD4}
U\sim 1+{C\over \sqrt{z}}~,
\end{equation}  
where $C$ is a positive integration constant.
This result is consistent with the $r\rightarrow \infty$ limit of 
$2Uz=rz^\prime$, which can be integrated to give the asymptotic behavior of 
the warp factors: $B(r)$ and $A(r)$ approach constant values as 
$r\rightarrow\infty$. Thus, these solutions are smooth and the 
volume is infinite.

{}On the other hand, the {\it positive} $\kappa$ solutions flow towards higher 
values of $U$ while $z$ decreases as $r\rightarrow \infty$. The numerical 
solution approaches the region $z\ll 1$, $U<-1$. By expanding $2zUf_2/f_1$ 
around $z=0$ we find the following asymptotic form of (\ref{dudz}):
\begin{equation}
-{16U^3\over U^2-1}{dU\over dz}\sim 1~,
\end{equation}
which has the following solution,
\begin{equation}
z\sim -8(U^2+{\rm ln}(U^2-1))+{\rm constant}~.
\end{equation}
Thus, as $z\rightarrow 0+$, $U\rightarrow U_*<-1$, a constant value. This in 
turn implies that in this limit ($r\rightarrow \infty$), $B(r)\sim 
(U_*-1){\rm ln}(r/r_0)$ and $A(r)$ approaches a constant value. 
It is important to note that for these solutions the volume is {\it 
finite}.

{}Let us now consider the possibility of interior solutions in which the volume
is automatically infinite. For these to 
satisfy (\ref{MC3}) with $z>0$, $U(\epsilon-)$ must lie in the region in 
between solid lines in Fig.5 (the dash-dotted line separates the positive 
$\kappa$ (right) region from the negative $\kappa$ (left) one). These 
solutions have potential finite distance singularities.
Indeed, with initial conditions in the allowed region, the solutions flow to 
negative values of $U$ and large values of $z$ approaching the curve $z=2U^2$ 
as $r$ decreases.
A large $U$ approximation of (\ref{dudr}) along this curve yields,
\begin{equation}
rU^\prime\sim U^2~.
\end{equation}
Thus, 
\begin{equation}\label{AD6}
U\sim -{1\over {\rm ln}(r/r_0)}~,
\end{equation}
where $r_0$ is an integration constant and $U$ diverges as $r\rightarrow r_0+$,
{\it i.e.}, we find a candidate for a finite distance singularity. 
However, since the asymptotic behavior of the warp factors implied 
by~(\ref{AD6}) is (up to overall numerical factors)
\begin{equation}
A(r)\sim B(r)\sim-{{\rm ln}({\rm ln}(r/r_0))}~,
\end{equation}
in the $r \rightarrow r_0+$ limit, this singularity is just a coordinate 
singularity; it is indeed not difficult to check that in this limit radial 
geodesics are in fact complete.

\begin{center}
{\it The $\zeta=-1$ Solutions}
\end{center}

{}Let us first study the exterior solutions. In order to have exterior 
solutions $U(\epsilon+)$ must lie to the right of
the solid line of Fig.6. In this region the string tension is negative. 
These solutions flow towards the curve  
$z=2U^2$ for large values of $U$, and along this curve to the leading order in 
$1/U$,
\begin{equation}
rU^\prime\sim U^2~.
\end{equation}
Thus,
\begin{equation}
U\sim{1\over {\rm ln}(r_0/r)}~,
\end{equation}
where $r_0$ is an integration constant. $U$ diverges in the limit $r 
\rightarrow r_0-$ and the asymptotic behavior of the warp factors in the same 
limit is 
as follows
\begin{equation}
A\sim B\sim -{\rm ln}({\rm ln}(r_0/r))~.
\end{equation}  
However, radial geodesics are complete in the limit $r\rightarrow r_0-$. And 
the volume of the extra-space is infinite.

{}Let us now consider the interior solutions.
With $z(\epsilon)>0$ and $U(\epsilon-)$ to the left of
 the solid line of Fig.6 we 
find solutions - with positive $\kappa$ - that flow towards large values of 
$z$ and $U\sim -1$ as $r$ 
decreases. In this region (\ref{dudz}) takes the asymptotic form
\begin{equation}
-{2z\over U+1}{dU\over dz}\sim 1~,
\end{equation}
which yields,
\begin{equation}
-2{\rm ln}\left|U+1\right|\sim {\rm ln}(z)+{\rm constant}~.
\end{equation}
This implies that $U\rightarrow -1$ as $z\rightarrow+\infty$. This is 
consistent with the $r\rightarrow 0$ limit; 
indeed, through $2zU=rz^\prime$ we obtain, 
in this asymptotic domain,
\begin{equation}
z\sim {1\over r^2}~.
\end{equation}
In this way we find the behavior of the warp factors in the $r\rightarrow 0$
limit:
\begin{eqnarray}
A&\sim& {\rm constant}~,\\
B&\sim& -2{\rm ln}(r/r_0)~.
\end{eqnarray}
These are {\it infinite} volume smooth solutions with {\it positive} $\kappa$.

{}Thus, as we see, in the case of a string in 5D bulk we have sensible
smooth infinite-volume solutions with positive brane tension if in the
bulk action we include both the Einstein-Hilbert and Gauss-Bonnet terms. 
Moreover, these solutions are {\em diluting}. 

\subsection{3-Brane in 7D Bulk}

{}In this subsection we study solutions with a flat 3-brane in 7D bulk space,
which is the most interesting case from the phenomenological point of view.
The ($rr$), ($\mu\nu$) and ($rr-\alpha\beta$) equations of motion respectively
read:
\begin{eqnarray}\label{EH3b1}
 &&6(A^\prime)^2+8N+S-12\xi\ {r^2\over \rho^2}\biggl\{(A^\prime)^4
 +4N^2+8(A^\prime)^2N+2 (A^\prime)^2S\biggr\}=0~,\\%[3mm]
 &&3(A^\prime)^2+3M+6N+2P+S-\nonumber\\
 &&12\xi\ {r^2\over\rho^2}\biggl\{2N^2
+(A^\prime)^2(M+2N+2P+S)+4MN+MS+2NP\biggr\}+
\nonumber\\&&{1\over 2}{\rm e}^{B}
\biggl[\t\Lambda-2\lambda\biggr]\t L\delta(r-\epsilon)=0~,\label{EH3b2}\\
%[3mm]
%
%
%
&&4(N-M)+S-P-24\xi\ {r^2\over \rho^2}\biggl\{2N(N-M)+(A^\prime)^2(2(N-M)+S-P)
\biggr\}-
\nonumber\\&&{1\over 2}{\rm e}^{B}
\t\Lambda\t L\delta(r-\epsilon)=0~.\label{EH3b3}
\end{eqnarray}
Using the notations of the previous subsection, the ($rr$) equation can 
be rewritten as 
\begin{equation}\label{EH3b4}
U^2(72V^2-z)-2U4V(z-12V^2)-z(6V^2-1)-12V^2(2-V^2)=0~,
\end{equation}
which implies 
\begin{eqnarray}\label{EH3b5}
U 
&=& {4V(z-12V^2)+\eta\sqrt{1440V^6-96zV^2+10z^2V^2+z^2+1728V^4
+60zV^4}\over 72V^2-z}~,
\end{eqnarray}
where\footnote{In the limit $\xi\rightarrow \infty$, where we can neglect 
the bulk Einstein-Hilbert term, these 
solutions correspond, respectively, to the 
$\eta=\pm 1$ 3-brane solutions found in the previous section.} $\eta=\pm 1$.  

{}Similarly, we can rewrite (\ref{EH3b3}) as follows
\begin{equation}
4\biggl(z-12V(U+V)\biggr)\biggl[(2U-V)V-rV^\prime\biggr]
+(z-24V^2)(U^2-1-rU^\prime)
={1\over 2\xi}\epsilon R\kappa\t L\delta(r-\epsilon)~. \label{EH3b6}
\end{equation}
Thus, away from the brane we obtain the following first order differential 
equation for $V$:
\begin{equation}\label{dvdr}
rV^\prime={g_1\over g_2}~,
\end{equation}
or treating $V$ as a function of $z$ we have:
\begin{equation}\label{dvdz}
2zU{dV\over dz}={g_1\over g_2}~,
\end{equation}
where
\begin{eqnarray}
g_1&=&8V^3z^2-192V^5z-4608V^5U^2+576V^6U-4320V^4U^3-z^2U^3+z^2U+2z^2V-
\nonumber\\&&
72zV^3+864V^4U+1152V^7+576V^5+672V^3zU^2-10Vz^2U^2-17V^2z^2U+
\nonumber\\&&
192V^4zU+108V^2U^3z-60zV^2U~,\nonumber\\
g_2&=&-12zVU^2+144zV^2U+108zV^3+12zV-5z^2V-864V^3U^2-1152V^4U\nonumber\\
&&-288V^3-864V^5~.\nonumber
\end{eqnarray}
To arrive this result we have replaced $U^\prime$ by the expression 
obtained by differentiating (\ref{EH3b4}) w.r.t. $r$ and we used the relation 
\begin{equation}\label{dzdr}
 rz^\prime=2zU~,
\end{equation}
which follows from the definitions of $U$ and $z$.

{}The full solutions must satisfy the jump conditions coming from (\ref{EH3b6})
and (\ref{EH3b2}). They read, respectively, 
\begin{eqnarray}
&&\left.\left[4(4V^2-z)V-zU+24V^2U\right]\right|^{\epsilon+}_{\epsilon-}=
{1\over 2\xi}R\kappa\t L~,\\
&&\left.\left[3Vz+2zU-4V^3-24V^2U-12V(U^2-1)\right]\right|^{\epsilon+}
_{\epsilon-}={1\over 2\xi}(2-\kappa)R\t L~.  
\end{eqnarray}
Let us rewrite these conditions as follows:
\begin{eqnarray}
&&\left.\left[12V^3+z(U-V)-12VU^2\right]\right|^{\epsilon+}_{\epsilon-}=
{R\t L\over \xi}~,\\
&&\left.\left[32V^3+48V^2U-2zU-8zV\right]\right|^{\epsilon+}
_{\epsilon-}={R\t L\over \xi}\kappa~.  
\end{eqnarray}
As in the previous section we will consider interior ($V(r>\epsilon)=0$, 
$U(r>\epsilon)=1$) as well as exterior ($V(r<\epsilon)=0$, $U(r<\epsilon)=1$) 
solutions. For these, the above matching conditions take the following form:
\begin{eqnarray}
&&\left.\left[12V^3+z(U-V-1)-12V(U^2-1)\right]\right|_{\epsilon\pm}=\pm
{R\t L\over \xi}~,\label{mc1}\\
&&\left.\left[32V^3+48V^2U-2z(U-1)-8zV\right]\right|_{\epsilon\pm}
=\pm{R\t L\over \xi}\kappa~,\label{mc2}  
\end{eqnarray}
where the plus sign corresponds to the exterior solutions while the minus 
sign corresponds to the interior ones. 

{}In order to have non-trivial 
solutions consistent with the matching conditions, 
$V(\epsilon+)$ (for exterior solutions), $V(\epsilon-)$ (for interior 
solutions) and $z(\epsilon)$ must be chosen in certain regions of 
the $V-z$ plane 
which are identified in Fig.7 and Fig.8 for the $\eta =+1$ and $\eta=-1$ 
cases, 
respectively. In what follows we will
study properties of the non-trivial parts of
the solutions with initial conditions ($V(\epsilon\pm),\ z(\epsilon)$) in the
allowed regions.

\begin{center}
{\it The $\eta=+1$ Solutions}
\end{center}

{}Let us first consider positive Gauss-Bonnet coupling, $\xi>0$ 
({\it i.e.}, $z>0$).
Exterior solutions are consistent with (\ref{mc1}) if $V(\epsilon+)$, 
$z(\epsilon)$ lie in regions $I$, 
$IV$ or $V$ of Fig.7 (through (\ref{mc2}) regions $I$ and $V$ are 
consistent with positive $\kappa$, while region $IV$ is consistent with 
negative $\kappa$).
The solutions with initial conditions in region $I$ flow to $-V\gg 1$, $z\gg 1$
where (\ref{dvdz}) becomes
\begin{equation}
a{dV\over dz}\sim {V\over z}~,
\end{equation}
where $a=3(4+\sqrt{10})/(5+14\sqrt{10})$. In this approximation,
\begin{equation}
z\sim C(-V)^a~,~~~~~~~U\sim -bV~,
\end{equation}
where $b=(4+\sqrt{10})/6$ and $C$ is an integration constant. Now we can
integrate (\ref{dzdr}) to obtain 
\begin{equation}
V\sim -{a\over 2b}{1\over {\rm ln}(r_0/r)}~,
\end{equation}
where $r_0>\epsilon$ is an integration constant and $V\rightarrow -\infty$ as 
$r\rightarrow r_0-$ (and $z\rightarrow+\infty$). Thus, the asymptotic behavior 
of the warp factors in the $r\rightarrow r_0-$ limit is as follows:
\begin{eqnarray}
A&\sim&{a\over 2b}{\rm ln}({\rm ln}(r_0/r))~,\\
{\rm exp}(B)&\sim&{{\widetilde C}\over r}
\left({\rm ln}(r_0/r)\right)^{-{a\over 2}}~,
\end{eqnarray} 
where ${\widetilde C}$ 
is a constant. It is not difficult to show that radial null 
geodesics terminate with finite affine parameter as $r\rightarrow r_0-$ and 
the curvature scalar diverges in this limit implying that these solutions 
have a finite distance singularity\footnote{This is the case for all the finite
distance singularities that we mention throughout this section.}. 

{}The solutions with initial 
conditions in regions $IV$ and $V$ flow to $V\gg 1$,
$z\gg 1$ approaching the curve $z=12V^2+6/5$ as $r$ increases. In this 
approximation $U\sim V$ and (\ref{dvdr}) can be integrated to give,
\begin{equation}
V\sim {1\over {\rm ln}(r_0/r)}~,
\end{equation} 
where $r_0>\epsilon$ is an integration constant and $V$ diverges as $r
\rightarrow r_0-$. 
In this limit the warp factors have the following asymptotic 
behavior:
\begin{eqnarray}
A&\sim&-{\rm ln}({\rm ln}(r_0/r))~,\\
{\rm exp}(B)&\sim&{{\widetilde C}\over r}\left({\rm ln}(r_0/r)\right)~,
\end{eqnarray} 
where ${\widetilde C}$ 
is an integration constant. As in the previous case, these 
solutions have finite distance singularities.

{}The 
interior solutions compatible with (\ref{mc1}) have initial conditions in 
regions $II$ and $III$ of Fig. 7 (the former is compatible with (\ref{mc2}) 
for $\kappa<0$ while the latter is compatible with $\kappa>0$).
The solutions with initial conditions such that $z(\epsilon)
{\ \lower-1.2pt\vbox{\hbox{\rlap{$<$}\lower5pt\vbox{\hbox{$\sim$}}}}\ }
24V^2(\epsilon-)$ within region $II$ flow to the origin of the $V-z$ plane. For
 $z\ll 1$, $-V\ll 1$ (\ref{dvdz}) becomes
\begin{equation}
2{dV\over dz}\sim {V\over z}~,
\end{equation} 
and in this approximation
\begin{equation}
U\sim U_*\equiv\sqrt{(24-a)/(72-a)}~,
\end{equation}
where $0<a<24$ is an integration constant. The behavior of the warp factors 
as $r\rightarrow 0+$ is:
\begin{eqnarray}
A&\sim&{\rm constant}~,\\
B&\sim&(U_*-1){\rm ln}(r/r_0)~,
\end{eqnarray}
where $r_0$ is a constant.

{}The solutions with initial conditions in the rest of region $II$ flow to
$-V\ll 1$, $z\gg 1$. In this approximation $U\sim -1$ 
and the warp factors have the
 following asymptotic behavior:
\begin{eqnarray}
A&\sim&{\rm constant}~,\\
B&\sim&-2{\rm ln}(r/r_0)~,
\end{eqnarray} 
where $r_0$ is an integration constant. The $r=0$ singularity here is a 
coordinate singularity, in fact, it is not difficult to show that 
radial geodesics are complete in the $r\rightarrow 0+$ 
limit.  

{}Some of the solutions with initial conditions in region $III$ flow to the 
region where the solutions become complex (these solutions are unphysical), 
some flow to the 
origin of the $V-z$ plane through region $V$, others flow to a finite value of 
$V=V_*>0$ as $z\rightarrow 0+$, 
and there are solutions that flow to $V\gg 1$, $z\gg 1$. 
For the latter solutions in the large $V$, large $z$ domain 
(\ref{dvdz}) becomes
\begin{equation}
a{dV\over dz}\sim {V\over z}~,
\end{equation}
where $a=3(4-\sqrt{10})\sqrt{10}/(14\sqrt{10}-5)$. In this approximation, 
$U\sim -bV$, where $b=(4-\sqrt{10})/6$, and we can integrate (\ref{dzdr}). 
We obtain:
\begin{equation}
V\sim {a\over 2b}{1\over{\rm ln}(r/r_0)}~,
\end{equation} 
where $r_0$ is an integration constant. 
$V$ diverges as $r\rightarrow r_0+$ and
the asymptotic behavior of the warp factors in this limit is given by:
\begin{eqnarray}
A&\sim&{a\over 2b}{\rm ln}({\rm ln}(r/r_0))~,\\
{\rm exp}(B)&\sim&{{\widetilde C}\over r}
\left({\rm ln}(r_0/r)\right)^{-{a\over 2}}~,
\end{eqnarray} 
where ${\widetilde C}$ is a constant. These solutions have a finite distance 
singularity.

{}The solutions that flow to $V_*\approx 0.25$ are also singular but in the 
$r\rightarrow 0$ limit. In this limit for these solutions the warp 
factors are given by: 
\begin{eqnarray}
A&\sim&-V_*{\rm ln}(r_0/r)~,\\
B&\sim&(1-U_*){\rm ln}(r/r_0)~,
\end{eqnarray}  
where $U_*\approx0.43$ and $r_0$ is an integration constant.

{}The solutions 
that flow to the origin are also singular in the $r\rightarrow 0$
limit.  The analysis for these solutions is similar to that for 
the interior solutions with initial 
conditions in region $II$ that flow to the origin.

{}Let us now consider the $\xi<0$ case ({\it i.e.}, $z<0$).  
Interior solutions are consistent with (\ref{mc1}) if ($V(\epsilon-)$, 
$z(\epsilon)$) lie in regions $I$ or $V$ of Fig. 7 which is consistent with 
(\ref{mc2}) if $\kappa>0$. The latter solutions have the same properties as the
interior solutions with initial conditions in region $III$ discussed above, 
while
the former ones flow to the origin of the $V-z$ plane and they are singular 
in the $r \rightarrow 0$ limit. The analysis is similar to that for 
the interior solutions of region $III$ with the only difference that the 
constant $a$ is negative in the present case.    

{}Exterior solutions are consistent with (\ref{mc1}) if ($V(\epsilon-)$, 
$z(\epsilon)$) lie in region $VI$ of Fig.7. If $z(\epsilon)
{\ \lower-1.2pt\vbox{\hbox{\rlap{$<$}\lower5pt\vbox{\hbox{$\sim$}}}}\ }
172.7V^2(\epsilon+)$ the solutions flow to $-z\gg 1$, 
$-V\ll 1$. In this domain (\ref{dvdz}) becomes
\begin{equation}
-{2\over \sqrt{10}}{dV\over dz}\sim {V\over z}~.
\end{equation}
In this approximation, $U\sim 1$, and thus $A$ and $B$ 
approach constant values in the large $r$ limit. Let us mention that these 
solutions are only consistent with negative values of $\kappa$. For initial 
conditions in the rest of region $VI$ the solutions have finite distance 
singularities. They flow to $-V\gg 1$, $-z\gg 1$ domain in which
\begin{equation}
a{dV\over dz}\sim {V\over z}~,
\end{equation}
where $a=2(4+\sqrt{10})/\sqrt{10}$. We obtain
\begin{equation}
V\sim -{6\over 10}{1\over {\rm ln}(r_0/r)}~,
\end{equation}
where $r_0$ is an integration constant. $V$ diverges as $r\rightarrow 
r_0-$ and the warp factors in this limit are given by:
\begin{eqnarray}
A&\sim&{6\over \sqrt{10}}{\rm ln}\left({\rm ln}(r_0/r)\right)~,\\
{\rm exp}(B)&\sim&{1\over r}\left({\rm ln}(r_0/r)\right)^{-{a\over 2}}~.
\end{eqnarray}  
It is not difficult to show that here we indeed have a singularity at $r=r_0$.

\begin{center}
{\it The $\eta=-1$ Solutions}
\end{center}

{}We will 
first focus on the positive Gauss-Bonnet coupling case, $\xi>0$  
({\it i.e.}, $z>0$).
Interior solutions are consistent with (\ref{mc1}) if ($V(\epsilon-)$, 
$z(\epsilon)$) lie in regions $I$, $II$, $V$ or $VI$ of Fig.8 
($I$ and $V$ are 
consistent with $\kappa>0$, while $II$ and $VI$ are consistent with 
$\kappa<0$).
The solutions from regions $I$ and $II$ flow to $z\gg 1$, $-V\gg 1$ approaching
the curve $z=12V^2$. In this domain (\ref{dvdr}) becomes
\begin{equation}
r{V^\prime}\sim V^2
\end{equation} 
with $U\sim V$. We obtain
\begin{equation}
V\sim -{1\over {\rm ln}(r/r_0)}~,
\end{equation}
where $r_0<\epsilon$ is an integration constant. $V$ diverges as $r\rightarrow
r_0+$ and in this limit the warp factors are given by:
\begin{eqnarray}
A&\sim&-{\rm ln}\left({\rm ln}(r/r_0)\right)~,\\
{\rm exp}(B)&\sim&{C\over r{\rm ln}(r/r_0)}~,
\end{eqnarray}  
where $C$ is a constant. The above expressions are divergent in the 
$r\rightarrow r_0+$ limit, but the singularity is just a coordinate 
singularity. Both the affine parameter of radial null geodesics and the proper 
time of time-like geodesics diverge in the same limit as $1/{\rm ln}(r/r_0)$. 
Thus, space is complete 
if we cut it at $r=r_0$. The values of $\kappa$ consistent with (\ref{mc2}) for
 initial conditions in region $I$ range from 0 along the dot-dashed curve   
growing to 2.15 for points away from it, while
 for initial conditions in region $II$ $\kappa$ ranges from $-\infty$ (for 
$z=12V^2$, $\kappa\sim 10V/3$ for $-V\gg 1$) to 0 (along the dot-dashed 
curve). The solutions with positive $\kappa$ are of particular interest and we 
will come to them later.

{}The solutions from regions $V$ and $VI$ flow to $z\gg 1$, $V\gg 1$ where 
(\ref{dvdz}) becomes 
\begin{equation}
a{dV\over dz}\sim {V\over z}~,
\end{equation}
where $a=3(4+\sqrt{10})\sqrt{10}/(5+14\sqrt{10})$. In this approximation,
\begin{equation}
z\sim CV^a~,~~~~~~~U\sim -bV~,
\end{equation}
where $b=(4+\sqrt{10})/6$ and $C$ is an integration constant. Now we can
integrate (\ref{dzdr}) to obtain 
\begin{equation}
V\sim {a\over 2b}{1\over {\rm ln}(r/r_0)}~,
\end{equation}
where $r_0<\epsilon$ is an integration constant and $V\rightarrow +\infty$ as 
$r\rightarrow r_0+$. The asymptotic behavior 
of the warp factors in the $r\rightarrow r_0+$ limit is as follows:
\begin{eqnarray}
A&\sim&{a\over 2b}{\rm ln}({\rm ln}(r/r_0))~,\\
{\rm exp}(B)&\sim&{{\t C}\over r}\left({\rm ln}(r/r_0)\right)^{-{a\over 2}}~,
\end{eqnarray} 
where ${\t C}$ is a constant. These solutions have a finite distance 
singularity.

{}Exterior solutions are consistent with (\ref{mc1}) if $V(\epsilon+)$, 
$z(\epsilon)$ lie in regions $III$ or $IV$. The former are consistent with
$\kappa>0$ while the latter are consistent with $\kappa<0$.
The positive $\kappa$ solutions flow to the regions where the solutions become 
complex (these solutions are unphysical), 
while the negative $\kappa$ solutions flow to $-V\ll 1$, $z\gg 1$. In
this domain
\begin{equation}
-2{dV\over dz}\sim {V\over z}~,
\end{equation}
which gives $V\sim C/z^2$ for constant $C$. In this approximation
$U\sim 1$, 
which allows us to integrate (\ref{dzdr}) to obtain the behavior of 
the warp factors: both $A$ and $B$ approach constant values in the $r
\rightarrow \infty$ limit. Let us mention that the volume for these smooth 
solutions is infinite. For these, the values of $\kappa$ range from 
approximately $-0.77$ to
0 (along the boundaries of the region).

{}Let us now study the negative Gauss-Bonnet coupling case ({\it i.e.}, $z<0$).
$V(\epsilon+)$, $z(\epsilon)$ in regions $I$ or $V$ of Fig.8 give exterior 
solutions consistent with (\ref{mc1}) and (\ref{mc2}) for 
positive values of $\kappa$. Both have finite distance singularities. Those 
from region $I$ flow to $-V\gg 1$, $-z\gg 1$ where (\ref{dvdz}) becomes 
\begin{equation}
a{dV\over dz}\sim {V\over z}~,
\end{equation}
where $a=3(\sqrt{10}-4)\sqrt{10}/(5-4\sqrt{10})$.  In this approximation,
\begin{equation}
z\sim C(-V)^a~,~~~~~~~U\sim -bV~,
\end{equation}
where $b=(4-\sqrt{10})/6$ and $C$ is an integration constant. We obtain 
\begin{equation}
V\sim -{a\over 2b}{1\over {\rm ln}(r_0/r)}~,
\end{equation}
where $r_0>\epsilon$ is an integration constant and $V\rightarrow -\infty$ as 
$r\rightarrow r_0-$. The asymptotic behavior 
of the warp factors in the $r\rightarrow r_0-$ limit is as follows:
\begin{eqnarray}
A&\sim&{a\over 2b}{\rm ln}({\rm ln}(r_0/r))~,\\
{\rm exp}(B)&\sim&{{\t C}\over r}\left({\rm ln}(r_0/r)\right)^{-{a\over 2}}~,
\end{eqnarray} 
where ${\t C}$ is a constant.

{}The solutions from region $V$ flow to the origin of the $V-z$ plane. For 
$V\ll 1$, $-z\ll 1$,
\begin{equation}
2{dV\over dz}\sim {V\over z}~.
\end{equation}  
Thus, $z\sim -aV^2$, where $a>0$ is an integration constant, and 
$U\sim U_*=\sqrt{(24+a)/(72+a)}$. Integrating (\ref{dzdr}) we obtain
\begin{equation}
V\sim-Cr^{-U_*}~,
\end{equation} 
where $C$ is an integration constant. $V$ diverges as $r\rightarrow 0+$. The
 warp factors are also divergent in this limit. 

{}Interior solutions are consistent with (\ref{mc1}) if $V(\epsilon-)$, 
$z(\epsilon)$ lie in region $VIII$ of Fig. 8. Throughout this region only 
negative values of $\kappa$ are consistent with (\ref{mc2}).
For $V(\epsilon-)<0$ the solutions flow to $-V\ll 1$, $-z\gg 1$ and in this 
domain we have $U\sim -1$. 
The behavior of the warp factors in the $r \rightarrow 
0$ limit is given by:
\begin{eqnarray}
A&\sim&{\rm constant}~,\\
B&\sim&-2{\rm ln}(r/r_0)~.
\end{eqnarray} 
The singularity in this case is just a coordinate singularity as geodesics are 
complete in the $r\rightarrow 0+$ limit.

{}For $V(\epsilon-)>0$ with 
$z(\epsilon)
{\ \lower-1.2pt\vbox{\hbox{\rlap{$<$}\lower5pt\vbox{\hbox{$\sim$}}}}\ }
-13.41V^2(\epsilon-)$ 
the solutions flow to $V\ll 1$, $-z\gg 1$ with the same asymptotic behavior of
the warp factors as in the previous case. 
For initial conditions in the rest of region $VIII$ the solutions flow to $V\ll
 1$, $-z\ll 1$ with the same behavior as for the interior solutions of region 
$V$ for the positive $z$ case. They have finite distance singularities.

\newpage
\begin{center}
{\it Diluting Solutions with Positive Tension}
\end{center}

{}Let us now come back to
the $\eta=-1$, $\xi>0$ interior solutions with initial 
conditions in region $I$ of Fig.8. From our discussion in the previous 
section, these solutions are of particular interest as some of these solutions 
have positive 3-brane tension (the ones with $\kappa>2$) and, furthermore, 
they are diluting.
 
{}We can 
for instance take the $z\ll |V|$ ($|V|\gg 1$) case in which (\ref{mc1}) and 
(\ref{mc2}) become, respectively,
\begin{eqnarray}
R{\widetilde L}/\xi&\approx& -11.77V^3(\epsilon-)~,\\
\kappa R{\widetilde L}/\xi&\approx& -25.3V^3(\epsilon-)~,
\end{eqnarray}
which gives $\kappa\approx 2.15$, and
$|V(\epsilon-)|\approx 10^{30}$, and $\epsilon$ is very close to the
would-be singularity $r_*$ (here for definiteness we have assumed the
extreme case $R^{-1}\sim {\widetilde M}_P\sim {\widehat M}_P$):
\begin{equation}
 {r_*\over \epsilon}-1\sim 10^{-30}~.
\end{equation}
Note that the singularity at $r=r_*>\epsilon$ would be there if we took the
interior solution for $r<\epsilon$ and continued it for values $r>\epsilon$.
However, in our solution (just as in the case without the Einstein-Hilbert 
term) there is no singularity as for $r>\epsilon$ the warp factors are actually
constant, and this is consistent with the matching conditions at $r=\epsilon$.
In particular, the solutions are
smooth everywhere, just as their counterparts from the previous section.
Thus, as we see, in the presence of both the 
Einstein-Hilbert and Gauss-Bonnet terms in the bulk action we have sensible
smooth solutions with positive brane tension. Moreover, these solutions are
{\em diluting}, that is, they exist for a range of values of the brane tension
(note that the Gauss-Bonnet coupling in these solutions is {\em positive}). 

{}Before we end this subsection, let us emphasize some important points. In the
diluting solutions we just discussed, we have a coordinate (but not a
real) singularity at some finite $r=r_0<\epsilon$. As we mentioned above, this
coordinate singularity is harmless as the corresponding geodesics are 
complete. Note that in the case without the Einstein-Hilbert term the 
corresponding coordinate singularity is at $r=0$. The reason why is the 
following. If we start with a solution corresponding to both 
the Einstein-Hilbert and Gauss-Bonnet terms present in the bulk, to arrive at 
the solution corresponding to only the Gauss-Bonnet term present in the bulk
we must take the limit $\xi\rightarrow\infty$, $M_P^{D-2}\xi={\rm
fixed}$. It is then not difficult to check that in this limit the coordinate
singularity at $r=r_0$ continuously moves to $r=0$. Also note that, since the
singularity at $r=r_0$ in the full solution is a coordinate singularity, we can
consistently cut the space at $r=r_0$. The geometry of the resulting solution
then is as follows. In the extra three dimensions we have a radially symmetric
solution where a 2-sphere is fibered over a semi-infinite line $[r_0,\infty)$.
The space is curved for $r_0\leq r<\epsilon$, at $r=\epsilon$ we have a jump
in the radial derivatives of the warp factors (because $r=\epsilon$ is where
the brane tension is localized), and for $r>\epsilon$ the space is actually
{\em flat}. So an outside observer located at $r>\epsilon$ thinks that the
brane is tensionless, while an observer inside of the sphere, that is, at
$r<\epsilon$ sees the space highly curved (and it would take this observer
infinite time to reach the coordinate singularity at $r=r_0$). This is an
important point, in particular, note that we did not find smooth exterior
solutions where the space would be curved outside but flat inside. And this is
just as well. Indeed, as was shown in \cite{Higher}, if we have only the
Einstein-Hilbert term in the bulk, then we have no smooth solutions whatsoever
(that is, smoothing out the 3-brane by making it into a 5-brane with two
dimensions curled up into a 2-sphere does not suffice). What is different in
our solutions is that we have added higher curvature terms in the bulk, which
we expected to smooth out some singularities. But serving as an ultra-violet
cut-off higher curvature terms could only possibly smooth out a real
singularity at $r<\epsilon$, not at $r>\epsilon$. And this is precisely what
happens in our solution - the presence of higher curvature terms ensures that
we have only a coordinate singularity at $r<\epsilon$ instead of a real naked
singularity as would be the case had we included only the Einstein-Hilbert 
term.  

\subsection{Implications for the Cosmological Constant Problem} 

{}In this section we saw that in the presence of the Einstein-Hilbert and
Gauss-Bonnet terms in the bulk action we have smooth infinite-volume solutions
which exist for a range of positive values of the brane tension (the diluting
property). These solutions, therefore, provide examples of brane world
scenarios where the brane world-volume can be flat without any fine-tuning
or presence of singularities. Is this then a solution to the cosmological
constant problem?

{}To answer this question, we need to address some issues here. First, note
that we have chosen a particular combination of quadratic in curvature
terms in the bulk action, namely, the Gauss-Bonnet combination. One could
argue that this is a fine-tuning as we have to fix 
two independent parameters at
the quadratic level to obtain the 
Gauss-Bonnet combination (note that one out of three {\em a priori} independent
parameters corresponds to the Gauss-Bonnet coupling, which is arbitrary in our
solutions). However, we suspect (albeit we do not have a proof of this 
statement) that even for generic higher curvature terms solutions with the
aforementioned properties should still exist. In particular, we suspect that 
the fact that we found non-singular solutions has to do with including
{\em higher curvature} terms in the bulk rather than with their particular
(Gauss-Bonnet) combination, which we have chosen to make computations 
tractable.

{}A more serious issue here has a phenomenological origin. Thus, as we 
discussed in subsection D of section III, the 6-dimensional Planck scale on the
5-brane (whose world-volume is a product of the 4-dimensional Minkowski 
space-time and a 2-sphere of a radius $R$) ${\widetilde
M}_P\gg M_P$ - as was explained in \cite{DGHS},
we must have 
the 7-dimensional bulk Planck scale $M_P\sim ({\rm millimeter})^{-1}$, so that 
the four-dimensional laws of gravity persist up the the distance scales of
order of the observable Hubble size. Naturally, here we can ask the following
question: Why is the 6-dimensional Planck scale on the 5-brane many 
(up to 30 in the extreme case where ${\widetilde M}_P\sim {\widehat M}_P$)
orders of 
magnitude larger than the 7-dimensional Planck scale? We would like to give one
speculative scenario for obtaining such a hierarchy of Planck scales. Thus,
let the 5-brane theory be a (non-conformal) large $N$ gauge theory. Then
the Planck scale induced on the brane due to quantum effects is expected to
be of order of ${\widetilde M}_P^4\sim N^2 M_P^4$. The required 
rank of the gauge theory in this case would have to be up to $N\sim 10^{60}$
(in the extreme case). It would be interesting to understand if one could
accommodate the Standard Model in such a scenario.

{}There are many interesting open questions to be addressed in scenarios with
infinite-volume extra dimensions. As was originally pointed out in 
\cite{DGP,witten,DG}, these scenarios offer a new arena for addressing the
cosmological constant problem. We hope our results we presented in this paper
are at least encouraging in this context. And addressing the 
aforementioned open questions definitely
seems to be worthwhile. In this light, we would
like to end this paper with the following poem by a 19th century Russian
poet Yuri Lermontov (translation from Russian by Z. Kakushadze): 

\begin{eqnarray}
&&{\mbox {\large{~~~~~~~~~~~The Sail}}}\nonumber\\
&&\phantom{}\nonumber\\
&&{\mbox {A lonely sail seeming white,}}\nonumber\\
&&{\mbox {In misty haze mid blue sea,}}\nonumber\\
&&{\mbox {Be foreign gale seeking might?}}\nonumber\\
&&{\mbox {Why home bays did it flee?}}\nonumber\\
&&\phantom{}\nonumber\\
&&{\mbox {The sail's bending mast is creaking,}}\nonumber\\
&&{\mbox {The wind and waves blast ahead,}}\nonumber\\
&&{\mbox {It isn't happiness it's seeking,}}\nonumber\\
&&{\mbox {Nor is it happiness it's fled!}}\nonumber\\
&&\phantom{}\nonumber\\
&&{\mbox {Beneath are running {\'a}zure streams,}}\nonumber\\
&&{\mbox {Above are shining golden beams,}}\nonumber\\
&&{\mbox {But wishing storms the sail seems,}}\nonumber\\
&&{\mbox {As if in storms is peace it deems.}}\nonumber
\end{eqnarray}

\acknowledgments

{}We would like to thank Gia Dvali and 
Gregory Gabadadze for valuable discussions.
O.C. is grateful to Max Rinaldi for fruitful discussions.
This work was supported in part by an Alfred P. Sloan Fellowship.

%%%%%%%%%%%%%%%%%%%%%%%%%%%%%%%%%%%%%%%%%%%%%%%%
\newpage
\begin{figure}[t]
%\hspace*{}
%\vspace*{}
\epsfxsize=14 cm
\epsfbox{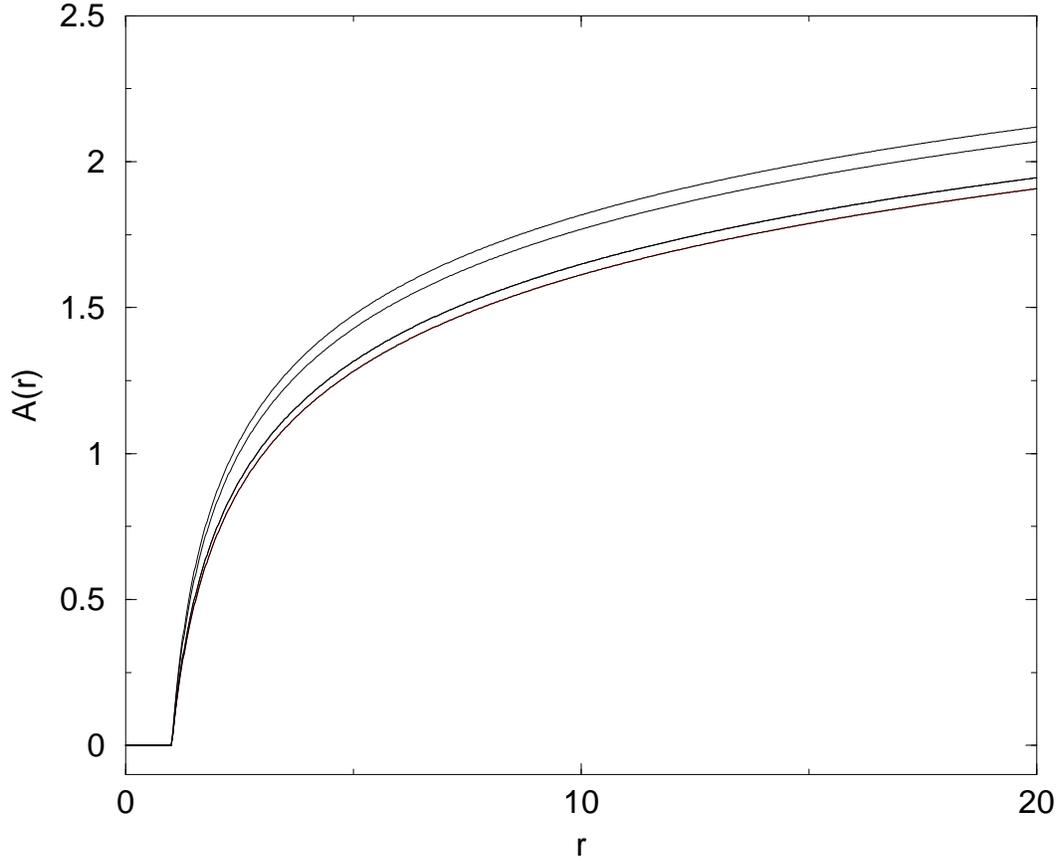}
\caption{Solutions for $A(r)$ (in $\epsilon=1$ units with $A(\epsilon)$ set to
zero) for various values of
$a\equiv\kappa{\widetilde L}$, $b\equiv(\kappa-2){\widetilde L}$ (and $\xi$ -
see below).
{}From top to bottom: 
$\xi=1$, $a=2$, $b=-1$; $\xi=1$, $a=1$, $b=-1$; 
$\xi=1$, $a=2$, $b=-1.2$; $\xi=1$, $a=1$, $b=-1.2$
(The $\xi=2$ counterparts of the lines for $B(r)$ in Fig.2 coincide with the
$\xi=1$ lines.) This plot shows that
Solution A for the membrane is indeed diluting.}
\end{figure}
%%%%%%%%%%%%%%%%%%%%%%%%%%%%%%%%%%%%%%%%%%%%%%%%

%%%%%%%%%%%%%%%%%%%%%%%%%%%%%%%%%%%%%%%%%%%%%%%%
\newpage
\begin{figure}[t]
%\hspace*{}
%\vspace*{}
\epsfxsize=14 cm
\epsfbox{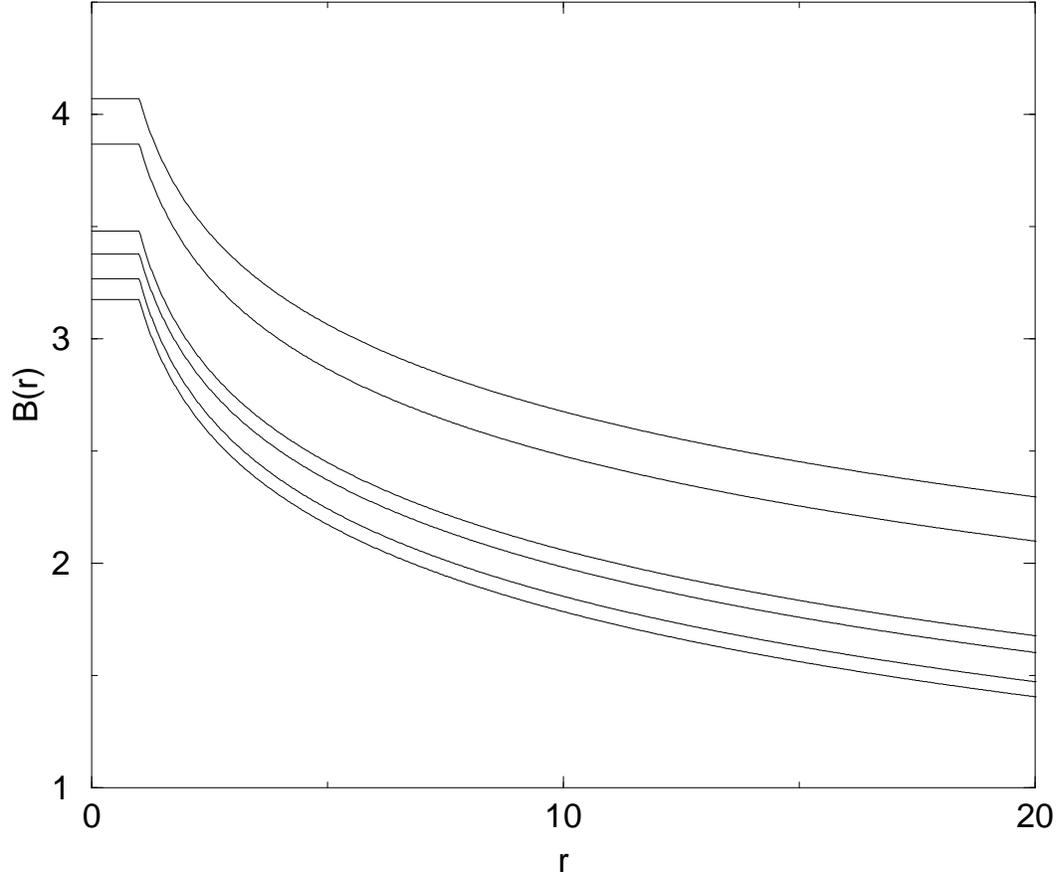}
\caption{Solutions for $B(r)$ (in $\epsilon=1$ units) for various values of
$a\equiv\kappa{\widetilde L}$, $b\equiv(\kappa-2){\widetilde L}$ and $\xi$.
{}From top to bottom: $\xi=2$, $a=1$, $b=-1$; $\xi=2$, $a=1$, $b=-1.2$;
$\xi=1$, $a=2$, $b=-1$; $\xi=1$, $a=1$, $b=-1$; 
$\xi=1$, $a=2$, $b=-1.2$; $\xi=1$, $a=1$, $b=-1.2$. This plot shows that
Solution A for the membrane is indeed diluting.}
\end{figure}
%%%%%%%%%%%%%%%%%%%%%%%%%%%%%%%%%%%%%%%%%%%%%%%%

%%%%%%%%%%%%%%%%%%%%%%%%%%%%%%%%%%%%%%%%%%%%%%%%
\newpage
\begin{figure}[t]
%\hspace*{}
%\vspace*{}
\epsfxsize=14 cm
\epsfbox{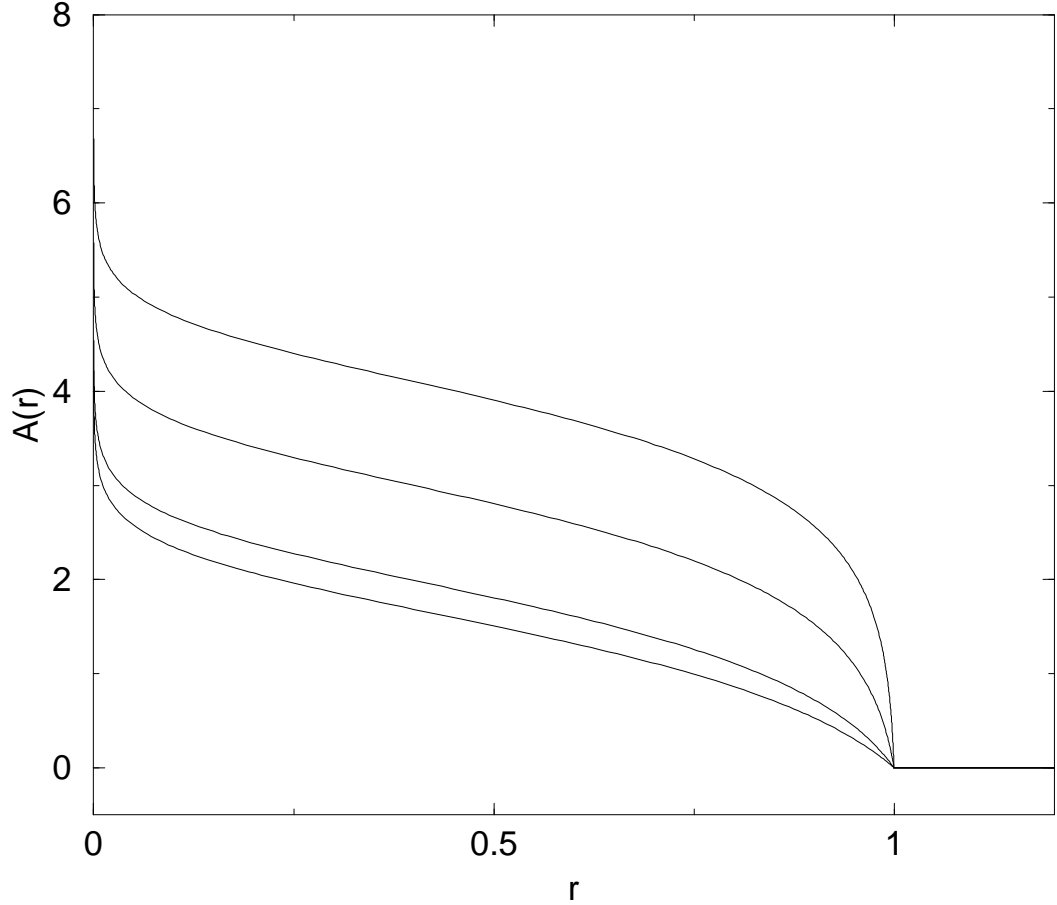}
\caption{Solutions for $A(r)$ (in $\epsilon=1$ units with $A(\epsilon)$ set to
zero) for ${\widetilde\Lambda}=1$ (see below) and various values of
$a\equiv {\widetilde L}/\xi$.
{}From top to bottom: 
$a=10^8$, $a=10^6$, $a=10^4$ and $a=1$.
(The ${\widetilde\Lambda}\not =1$ counterparts of the solid lines for $B(r)$ 
in Fig.4 cannot be distinguished from the
$a={\widetilde\Lambda}=1$ line.) This plot shows that the $\xi>0$ branch of 
the $\eta=-1$ 3-brane solution is indeed diluting.}
\end{figure}
%%%%%%%%%%%%%%%%%%%%%%%%%%%%%%%%%%%%%%%%%%%%%%%%

%%%%%%%%%%%%%%%%%%%%%%%%%%%%%%%%%%%%%%%%%%%%%%%%
\newpage
\begin{figure}[t]
%\hspace*{}
%\vspace*{}
\epsfxsize=14cm
\epsfbox{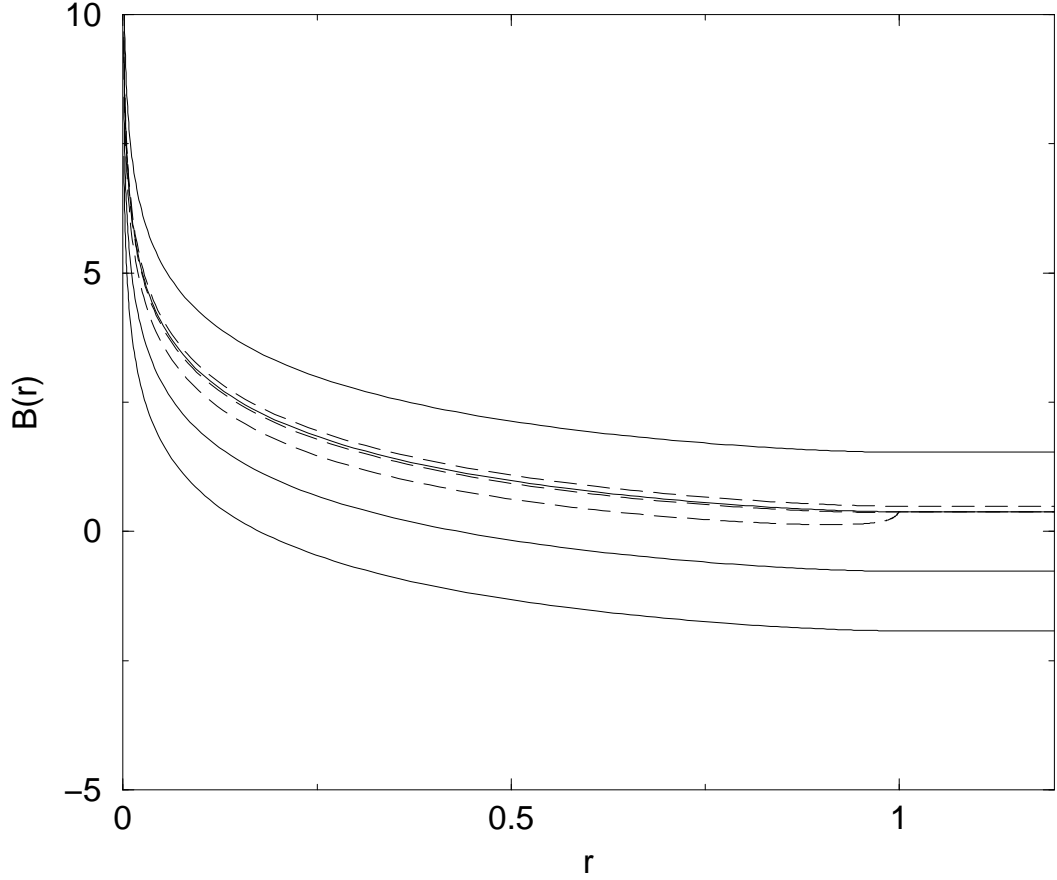}
\vspace{1cm}
\caption{Solutions for $B(r)$ (in $\epsilon=1$ units) for various values of
${\widetilde\Lambda}$ and $a\equiv{\widetilde L}/\xi$.
{}From top to bottom: $a=1$ with ${\widetilde\Lambda}=.1$, 
${\widetilde\Lambda}=1$, ${\widetilde\Lambda}=10$, ${\widetilde\Lambda}=100$
(solid lines) ; 
and ${\widetilde\Lambda}=1$ with $a=10^{-6}$, $a=10^4$, $a=10^8$ 
(dashed lines) ; 
This plot shows that the $\xi>0$ branch of the $\eta=-1$  
3-brane solution is indeed diluting.}
\end{figure}
%%%%%%%%%%%%%%%%%%%%%%%%%%%%%%%%%%%%%%%%%%%%%%%%

%%%%%%%%%%%%%%%%%%%%%%%%%%%%%%%%%%%%%%%%%%%%%%%%
\newpage
\begin{figure}[t]
%\hspace*{}
%\vspace*{}
\epsfxsize=14cm
\epsfbox{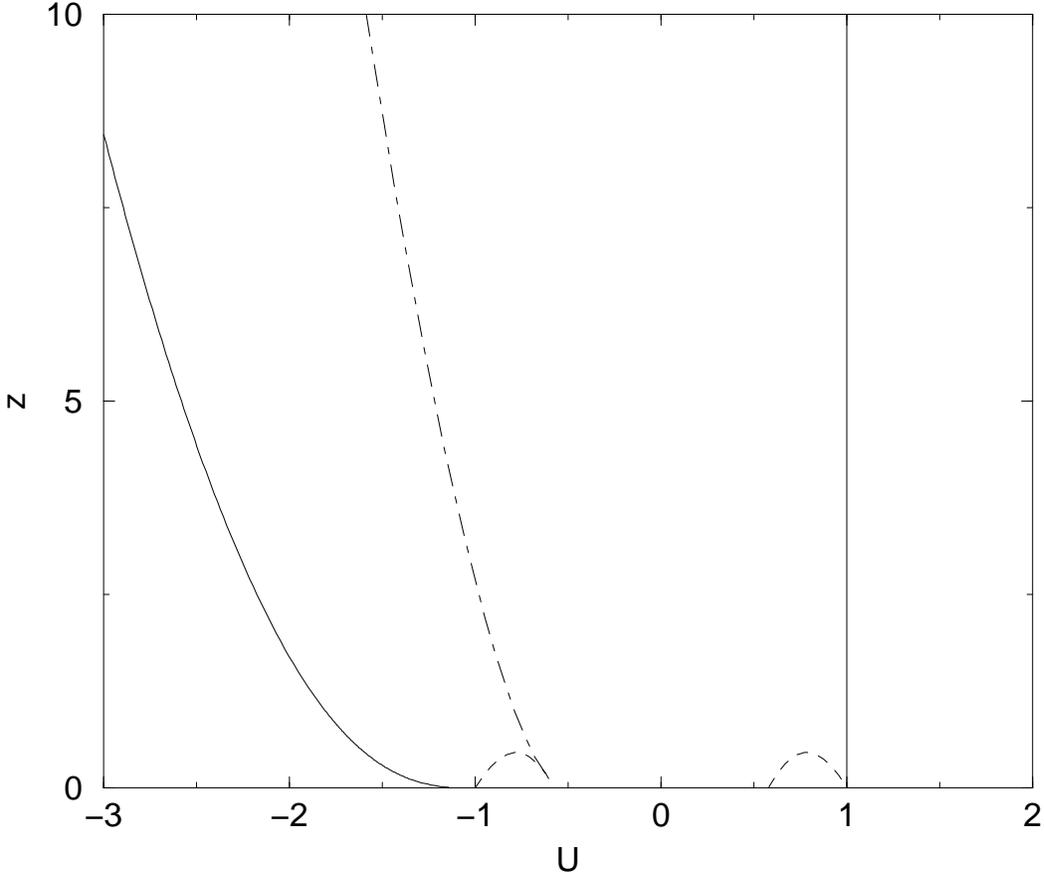}
\vspace{1cm}
\caption{The solid lines are zeros of $4VU(V+U)+4V+z(U-V-1)$ while the 
dot-dashed lines are the zeros of $-2z(2V+U-1)+8V^2U$. In the region in between
dashed lines there are no real solutions of (\ref{AD1}).}
\end{figure}
%%%%%%%%%%%%%%%%%%%%%%%%%%%%%%%%%%%%%%%%%%%%%%%%

%%%%%%%%%%%%%%%%%%%%%%%%%%%%%%%%%%%%%%%%%%%%%%%%
\newpage
\begin{figure}[t]
%\hspace*{}
%\vspace*{}
\epsfxsize=14cm
\epsfbox{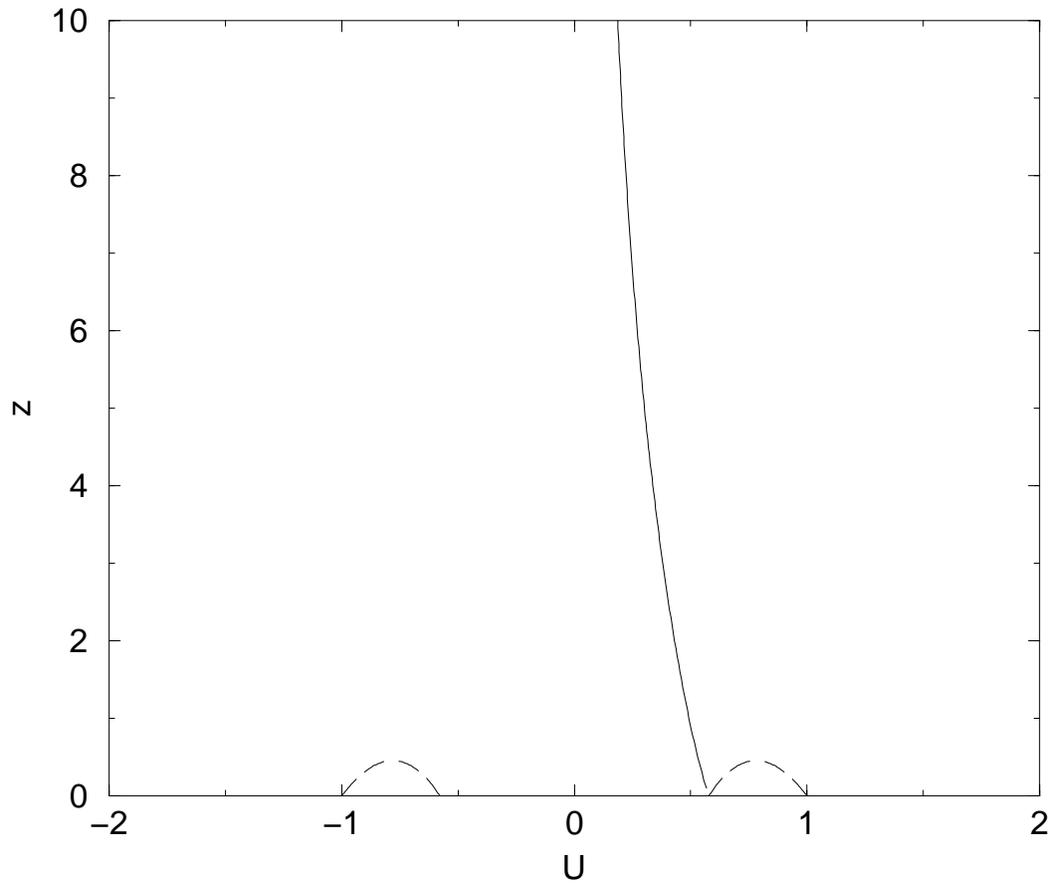}
\vspace{1cm}
\caption{The solid line represents 
the zeros of $4VU(V+U)+4V+z(U-V-1)$; inside the 
dashed pockets there are no real solutions of (\ref{AD1}).}
\end{figure}
%%%%%%%%%%%%%%%%%%%%%%%%%%%%%%%%%%%%%%%%%%%%%%%%

%%%%%%%%%%%%%%%%%%%%%%%%%%%%%%%%%%%%%%%%%%%%%%%%
\newpage
\begin{figure}[t]
%\hspace*{}
%\vspace*{}
\epsfxsize=14cm
\epsfbox{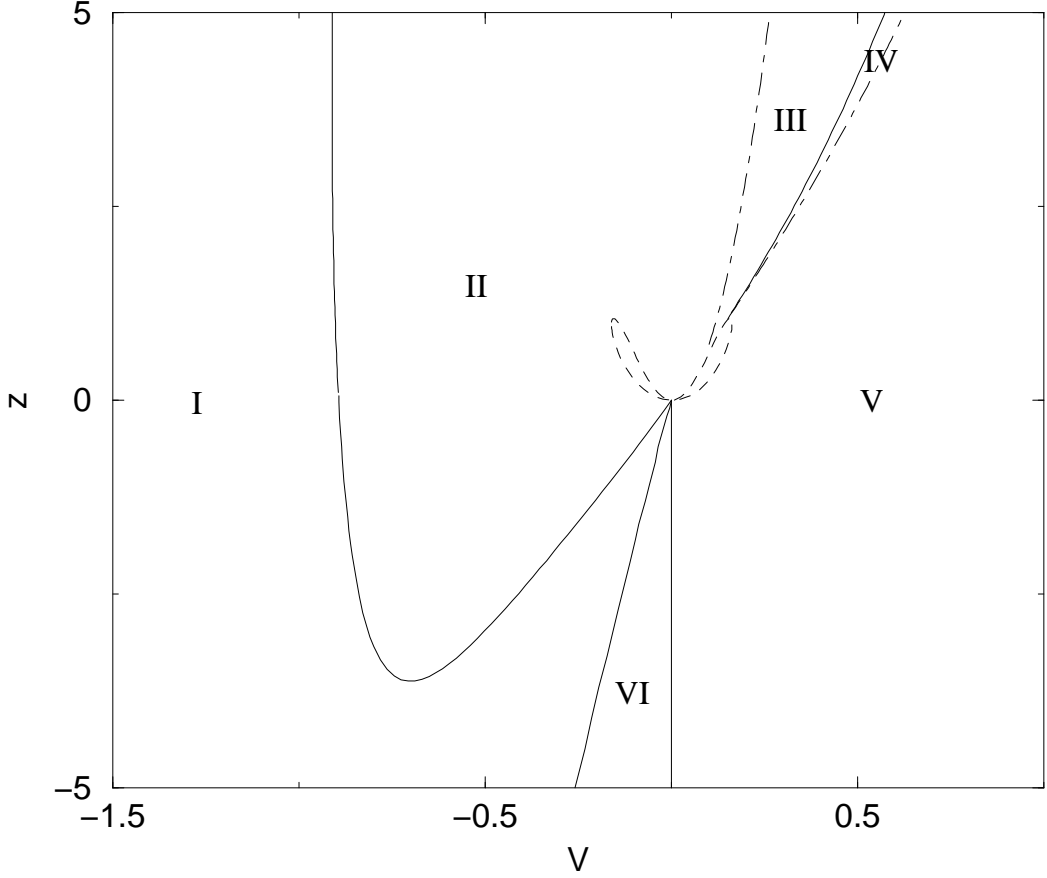}
\vspace{1cm}
\caption{The solid lines are zeros of $12V^3+z(U-V-1)-12V(U^2-1)$ that approach
the following curves for $\left|z\right|\gg 1$: $I-II$ boundary: $V=-2/3$, $III
-IV$ boundary: $z=12V^2$ and $I-VI$ boundary: $z=-27.3V^2$. The dot-dashed line
between $II$ and $III$ is $z=72V^2$ while the one between $IV$ and $V$ that 
approaches $z=9.6V^2$ for $z\gg 1$ consists of zeros of $\kappa$.  Inside 
the dashed pockets there are no real solutions of (\ref{EH3b4}).}
\end{figure}
%%%%%%%%%%%%%%%%%%%%%%%%%%%%%%%%%%%%%%%%%%%%%%%%

%%%%%%%%%%%%%%%%%%%%%%%%%%%%%%%%%%%%%%%%%%%%%%%%
\newpage
\begin{figure}[t]
%\hspace*{}
%\vspace*{}
\epsfxsize=14cm
\epsfbox{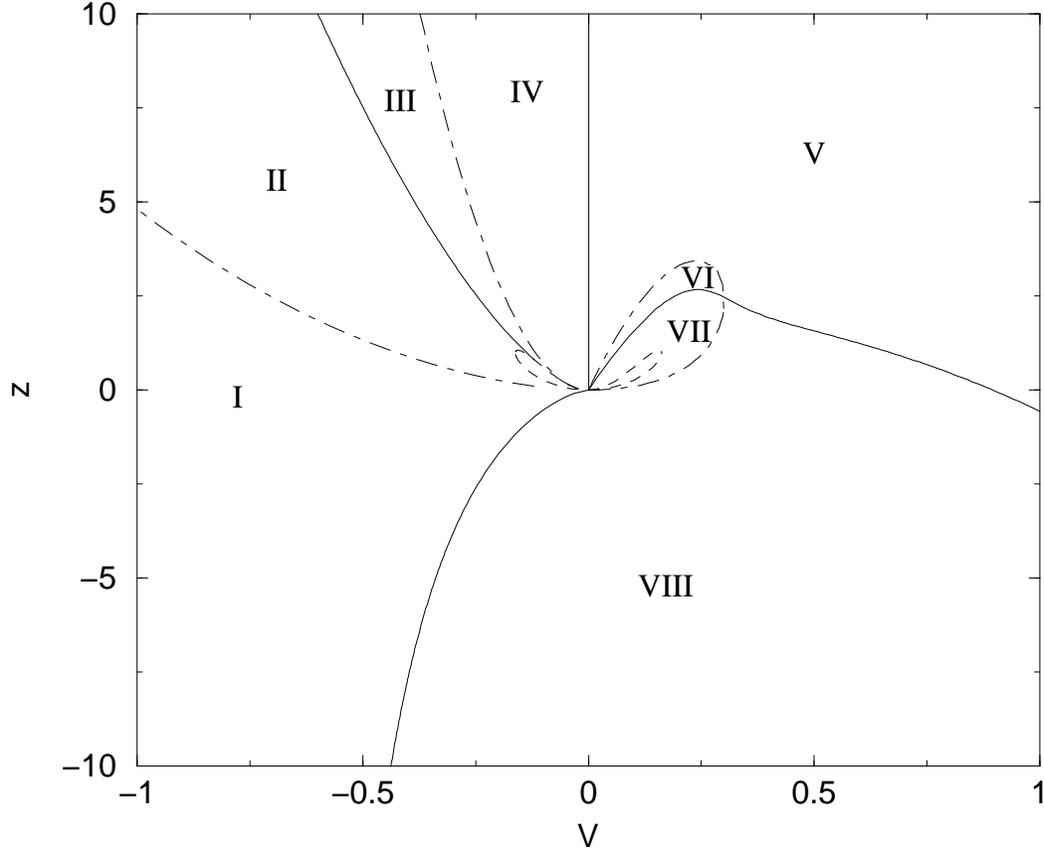}
\vspace{1cm}
\caption{The solid lines are the zeros of $12V^3+z(U-V-1)-12V(U^2-1)$ that 
approach the following curves for $\left|z\right|\gg 1$: $II-III$ boundary:
$z=12.7V^2$, $V-VIII$ boundary: $z=-3.66V^2$ and $I-VIII$ boundary: V=-2/3.
The dot-dashed line between $III$ and $IV$ is $z=72V^2$ while the one between 
$I$ and $II$ representing the zeros of $\kappa$ 
approaches $z=-3.66V^2$ for $V\gg 1$ (the 
third dot-dashed line consists also of zeros of $\kappa$). Inside 
the dashed pockets there are no real solutions of (\ref{EH3b4}).}
\end{figure}
%%%%%%%%%%%%%%%%%%%%%%%%%%%%%%%%%%%%%%%%%%%%%%%%

\end{document}